\documentclass[useAMS,usenatbib]{mn2e}

\usepackage{verbatim,graphicx,epsfig,dcolumn}
\newcolumntype{.}{D{.}{.}{4}}
\newcolumntype{,}{D{.}{.}{2}}
\newcolumntype{;}{D{.}{.}{1}}
\newcommand{\nodata}{$\cdot\cdot\cdot$}
\newcommand{\lesssim}{{\lower-1.2pt\vbox{\hbox{\rlap{$<$}\lower5pt\vbox{\hbox{$\sim$}}}}}}
\newcommand{\gtrsim}{{\lower-1.2pt\vbox{\hbox{\rlap{$>$}\lower5pt\vbox{\hbox{$\sim$}}}}}}

\title[VISTA variables in the Sgr dSph]{VISTA variables in the Sagittarius dwarf spheroidal galaxy: pulsation- vs.\ dust-driven winds on the giant branches}
\author[I. McDonald, et al.]{I.~McDonald$^{1}$\thanks{E-mail: iain.mcdonald-2@manchester.ac.uk}, A.~A.~Zijlstra$^{1}$, G.~C.~Sloan$^{2}$, E.~Kerins$^{1}$, E.~Lagadec$^{2}$, D.~Minniti$^{3}$\\
$^{1}$Jodrell Bank Centre for Astrophysics, Alan Turing Building, Manchester, M13 9PL, UK\\
$^{2}$Cornell University, Astronomy Department, Ithaca, NY 14853-6801, USA\\
$^{3}$Departamento de Astronomia y Astrofisica, Pontificia Universidad Cat\'olica de Chile, Vicu\~na Mackenna 4860, Casilla 306,\\ \, Santiago 22, Chile 
}

\begin{document}

\date{Accepted 9999 December 32. Received 9999 December 32; in original form 9999 December 32}

\pagerange{\pageref{firstpage}--\pageref{lastpage}} \pubyear{9999}

\maketitle

\label{firstpage}

\begin{abstract}
Variability is examined in over 2.6 million stars covering 11 square degrees of the core of the Sagittarius dwarf spheroidal galaxy (Sgr dSph) from VISTA $Z$-band observations. Generally, pulsation on the Sgr dSph giant branches appears to be excited by the internal $\kappa$ mechanism. Pulsation amplitudes appear identical between red and asymptotic (RGB/AGB) giant stars, and between unreddened carbon and oxygen-rich stars at the same luminosity. The lack of correlation between infrared excess and variability among oxygen-rich stars indicates pulsations do not contribute significantly to wind driving in oxygen-rich stars in the Sgr dSph, though the low amplitudes of these stars mean this may not apply elsewhere. The dust-enshrouded carbon stars have the highest amplitudes of the stars we observe. Only in these stars does an external $\kappa$-mechanism-driven pulsation seem likely, caused by variations in their more-opaque carbon-rich molecules or dust. This may allow pulsation driving of winds to be effective in carbon stars. Variability can be simplified to a power law ($A \propto L / T^2$), as in other systems. In total, we identify 3\,026 variable stars (with r.m.s.\ variability of $\delta Z \gtrsim 0.015$ magnitudes), of which 176 are long-period variables associable with the upper giant branches of the Sgr dSph. We also identify 324 candidate RR Lyrae variables in the the Sgr dSph and 340 in the outer Galactic Bulge.
\end{abstract}



\section{Introduction}
\label{IntroSect}

The death of low- and intermediate-mass stars (0.8--8 M$_\odot$) occurs through stellar mass loss provoked by a pulsation-enhanced, radiation-driven wind. This process ejects the stellar atmosphere, quenching stellar fusion. There is a complex interaction between the driving of mass loss and the underlying stellar evolution. The timing and efficiency of this mass loss can dictate the chemistry and mineralogy of the material returned by the star to the host system. The integrated effect of this return of matter is therefore important in determining the elemental and molecular abundances, dust-to-gas ratios and mineralogy of the interstellar media of galaxies. Understanding how this process varies throughout Universal history is vital for a complete chemical view of the Universal elemental enrichment.

Despite recent advances, it is unclear what determines when and how this wind starts. It is thought that stars on the asymptotic giant branch (AGB) become unstable to pulsation, probably through the unstable ionisation of hydrogen via the $\kappa$-mechanism, which provokes a Cepheid-like opacity modulation in the star (e.g.\ \citealt{OW05}). These pulsations become strong enough to levitate atmospheric material, which cools until it can form molecules and then dust. Radiation pressure on these dust grains accelerates them away from the star, and collisional coupling with the surrounding gas means both media are ejected (e.g.\ \citealt{Kwok75}; \citealt{Bowen88}; \citealt{HO03}).

AGB stars tend to come in two flavours, depending on the surface ratio of carbon to oxygen. Stars typically begin the AGB as oxygen-rich stars, then progressively dredge up carbon from their interiors via a series of thermal pulses, triggered by further instabilities between the hydrogen- and helium-burning shell sources. If the C/O ratio exceeds unity, the star becomes a carbon star. Only certain stars will become carbon-rich, dictated by their initial mass and metallicity (e.g.\ \citealt{MWZ+12}, their figure 9, after \citealt{MG07}). For carbon-rich stars, pulsation-enhanced, radiation-driven mass loss appears to be effective at driving a stellar wind, as carbon grains are highly opaque \citep{LZ08}.

For oxygen-rich stars, particularly those in metal-poor environments, opacity-driven winds appear less successful. The most prevalent form of oxygen-rich dust, iron-poor amorphous silicate, is too transparent at the near-infrared peak of the star's luminosity to drive a wind in this manner \citep{Woitke06b}. In metal-poor stars, the decreased dust-to-gas ratio means each dust grain must share its momentum with more gas. In dense environments, where the mean free path of dust grains is too short, this added drag force may prevent the wind from escaping. In tenuous winds, where the mean free path is long, the dust can drift ahead of the gas at considerable velocity: the gas can fail to be ejected and fall back, and the dust grains can be destroyed by sputtering in collisions with the gas \citep{Kwok75}.

Several possibilities exist for solving the problem of driving metal-poor, oxygen-rich winds. Typical solutions include providing an additional energy input to the wind, or adding opacity to the dust. The most obvious source for additional energy input is increasing the radial range over which pulsations are an effective accelerant, usually taking advantage of turbulent motions or shocks (e.g.\ \citealt{Woitke06a}). The dust opacity can be increased by changing either the dust species or the grain properties. High-opacity dust species include amorphous carbon, theoretically allowed via CO dissociation but largely unobserved in oxygen-rich environments \citep{HA07}, or metallic iron \citep{KdKW+02,VvdZH+09,MSZ+10}. Alternatively, larger silicate grains scatter radiation more efficiently \citep{Hoefner08} and have been detected \citep{NTI+12}. All three dust species produce featureless infrared excess, and are thus indistinguishable in infrared spectra. A complete absence of silicate features in metal-poor stars ([Fe/H] $\lesssim$ --1 dex; \citealt{MSZ+10}), and a likely change in silicate composition at low metallicity \citep{JKS+12}, suggest some change in dust composition or properties with metallicity exists. Even with the increased opacity of these other dust species, it is not clear that radiation pressure alone can overcome stellar gravity \citep{MBvLZ11,MvLS+11,MZRJ13}.

In principle, differentiating between further energy input from pulsation, or `exotic' dust production is simple. If pulsation enhances the wind further, then mass-loss rate should scale with the amount of potential energy stored in the pulsations, which correlates with their amplitude \citet{vLMC+06}. The mass-loss rates of otherwise-identical stars at different metallicities should then be similar (subject to the remaining influence of radiation pressure), as should those of carbon- and oxygen-rich stars. If a different dust species provides opacity and radiation pressure still dominates the energy input, then the stellar wind velocity should scale with stellar luminosity and metallicity.

In practice, differentiating the two scenarios is complex, as pulsation energy and luminosity are also correlated. We must therefore look to see how parameters which are not directly affected by pulsation or luminosity affect the mass-loss rate, such as metallicity, C/O ratio and stellar mass. The mass-loss rate itself must be examined through proxies such as infrared excess. Our previous analyses of the \emph{Hipparcos} stellar sample \citep{MZB12} and globular cluster stars \citep{MBvLZ11,MvLS+11} find mass-losing stars are generally strongly pulsating, confirming the need for stellar pulsations. However, they lack the range in mass and chemistry required to make concrete statements about the overall role of pulsation in driving the wind. If pulsation drives the wind, we can expect that infrared excess should correlate with pulsation energy. If dust drives the wind, we can expect that infrared excess correlates with metallicity and C/O ratio instead.

We herein turn our attention to the Sagittarius dwarf spheroidal (Sgr dSph). At 25 kpc \citep{MKS+95,MBFP04,KC09}, it is our closest galactic neighbour\footnote{Ignoring the disputed Canis Major dwarf spheroidal, see \citet{LCMZ+12}.}. Although being tidally destroyed by the Milky Way, it retains a sizeable core that extends for several degrees, with its northern extent hidden behind the Galactic Bulge \citep{IGI94}. Although containing several populations \citep{SDM+07}, its core is strongly dominated by a population with --0.7 $\lesssim$ [Fe/H] $\lesssim$ --0.4 dex and an age of 8 $\pm$ 1.5 Gyr \citep{BCF+06}. The exception is the globular cluster M54, which lies on the projected centre of the galaxy and, though metal-poor ([Fe/H] = --1.7 dex, $t$ $\approx$ 13 Gyr; \citealt{SDM+07}), probably acts as the galaxy's core. We recently showed the bulk population to produce the majority of the Sgr dSph's carbon stars, despite its stars only just living long enough to make the carbon-rich transition \citep{MWZ+12}.

This work builds on the photometric survey of the Sgr dSph at $Z$, $J$ and $K_{\rm s}$ from the Visible and Infrared Survey Telescope for Astronomy (VISTA) (published as \citealt{MZS+13}, hereafter Paper II). With this ten-epoch dataset, we explore the variability of stars in the inner regions of the Sgr dSph, how it scales with various stellar parameters and with the infrared excess indicative of dust production. We combine it with our earlier Very Large Telesecope Fibre Large Array Multi-Element Spectrograph (VLT/FLAMES) spectroscopic survey (\citealt{MWZ+12}, hereafter Paper I) to determine differences in variability between carbon- and oxygen-rich stars. This work exists as part of a larger survey to address the properties of late-stage stellar evolution and mass loss at a range of stellar masses and metallicities (e.g.\ \citealt{MvL07,vLBM08,MvLD+09,BMvL+09,MvLS+11,MJZ11,BSvL+11,SML+12}), and complements work by others in environments with mass and metallicity ranges far removed from that of the Milky Way (e.g.\ \citealt{ITM+04,WMF08,FCRC+10,IGFH11,GCC+13,TSI13}).

\section{Observations and analysis}
\label{ObsSect}

\subsection{Calculating a variability index}
\label{CalcVarSect}

Details of the VISTA photometric observations can be found in Paper II. Briefly summarised, the observations cover seven tiles, each of which is 1.5 $\times$ 1.0 square degrees, placed on the known over-densities of Sgr dSph stars, in an irregular polygon around M54. Each tile was observed in $Z$ over either 12 or 13 epochs taken between 2012 April 06 and 2012 August 01 UT. Through tile overlaps, a small subset of objects are covered in up to 48 epochs. Tile-by-tile data catalogues were automatically reduced with the VISTA version 1.2 pipeline at the Cambridge Astronomy Survey Unit (CASU).


\begin{figure}
\centerline{\includegraphics[height=0.47\textwidth,angle=-90]{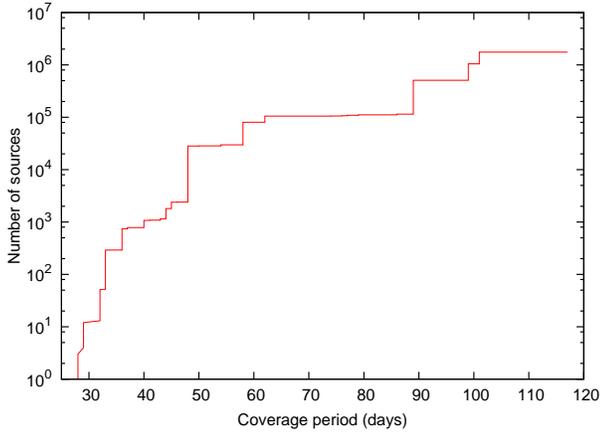}}
\caption{Number of sources which have at least the indicated period of coverage in our data.}
\label{TimebaseFig}
\end{figure}

\begin{figure}
\centerline{\includegraphics[height=0.47\textwidth,angle=-90]{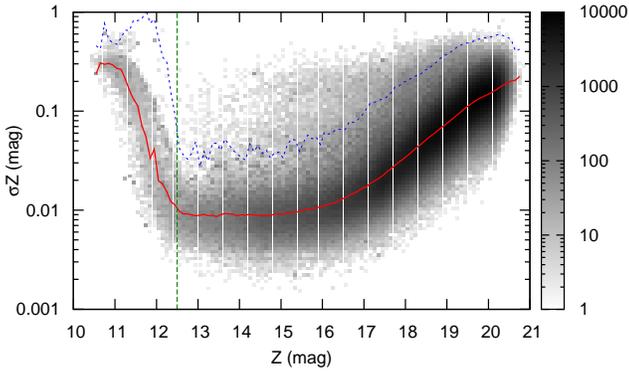}}
\caption{Magnitude--variability diagram of VISTA data. Sources brighter than $Z = 12.5$ mag are largely saturated, thus we have placed a limit (green, vertical line) beyond which we do not consider sources to be variable. The variability locus, representing the median variability per bin, is shown as a solid, red line. The cut-off for identifying variability is shown as the upper, blue, dashed line.}
\label{MagVarFig}
\end{figure}

\begin{figure*}
\centerline{\includegraphics[height=0.95\textwidth,angle=-90]{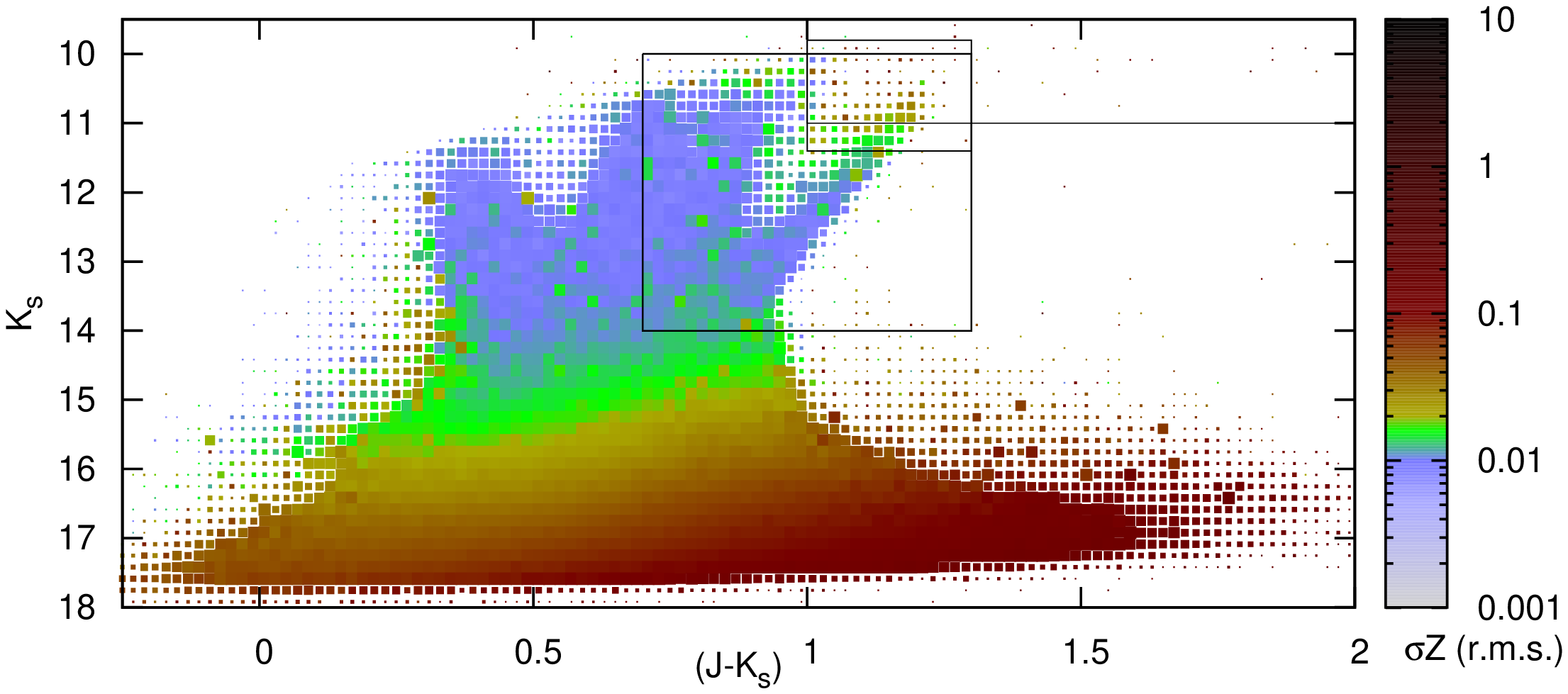}}
\centerline{\includegraphics[height=0.45\textwidth,angle=-90]{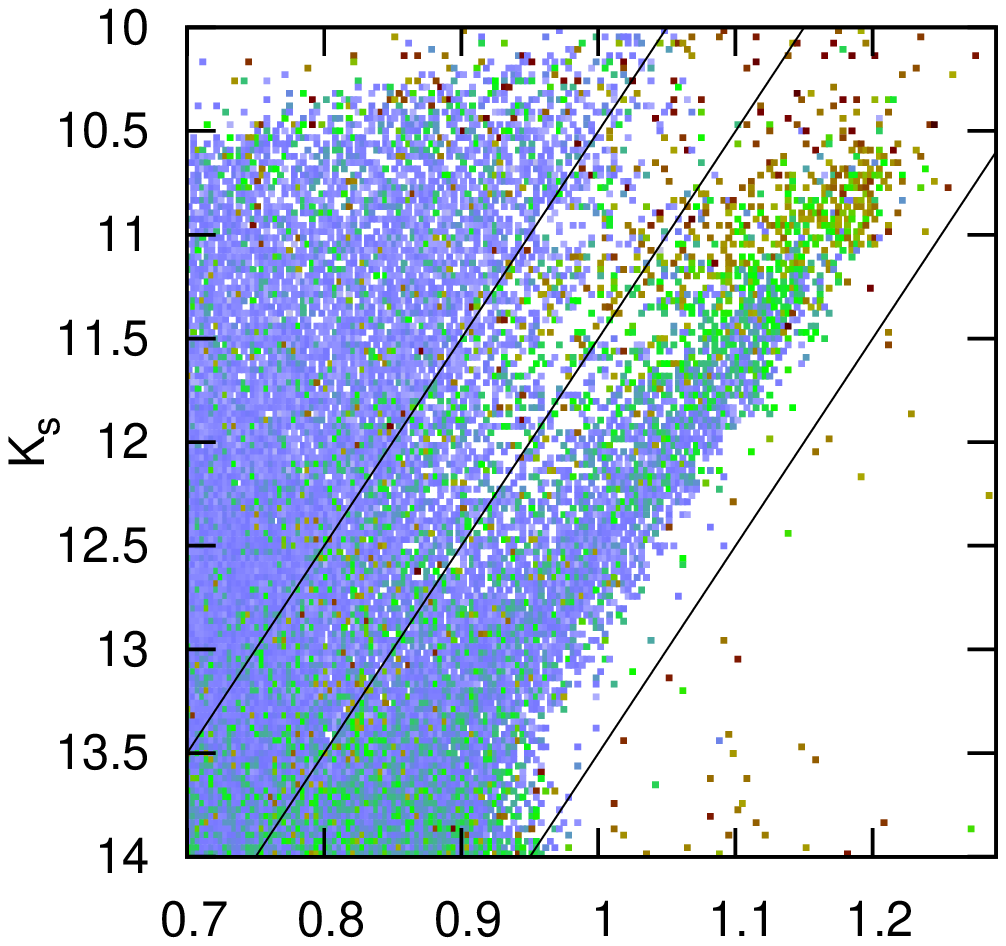}
            \includegraphics[height=0.45\textwidth,angle=-90]{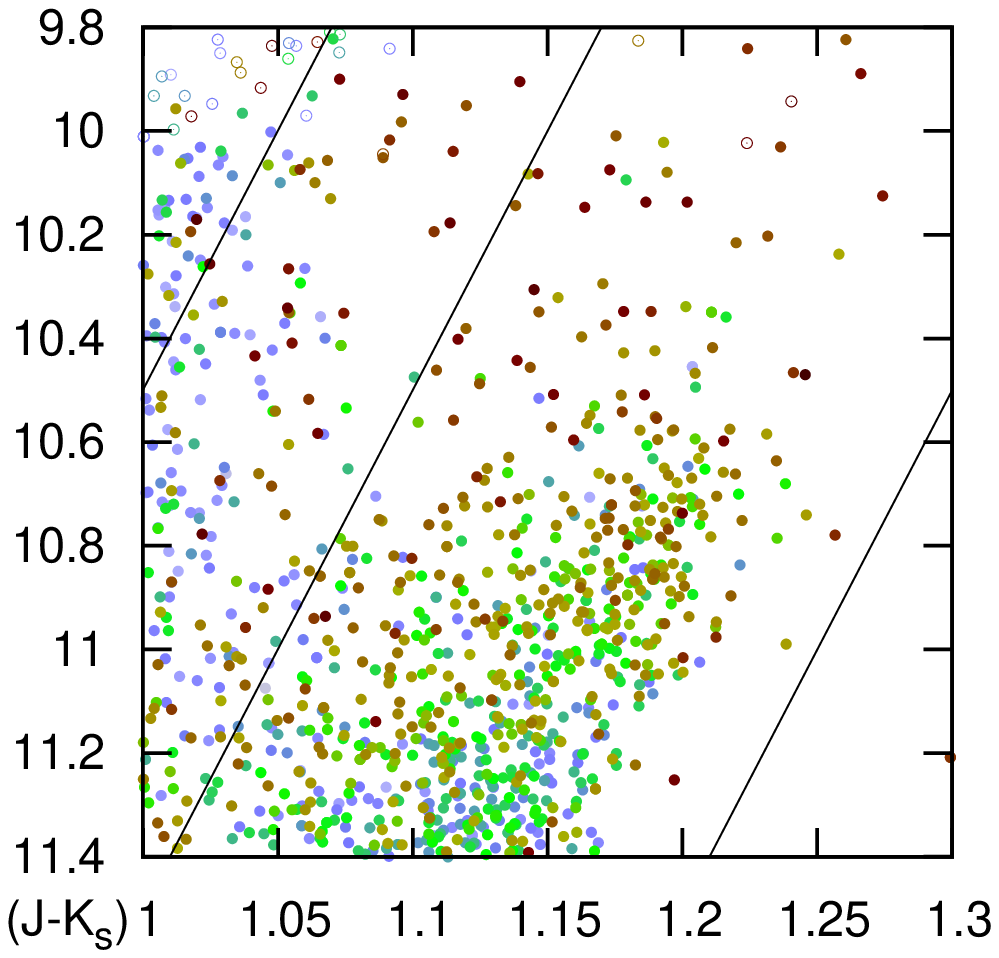}}
\centerline{\includegraphics[height=0.95\textwidth,angle=-90]{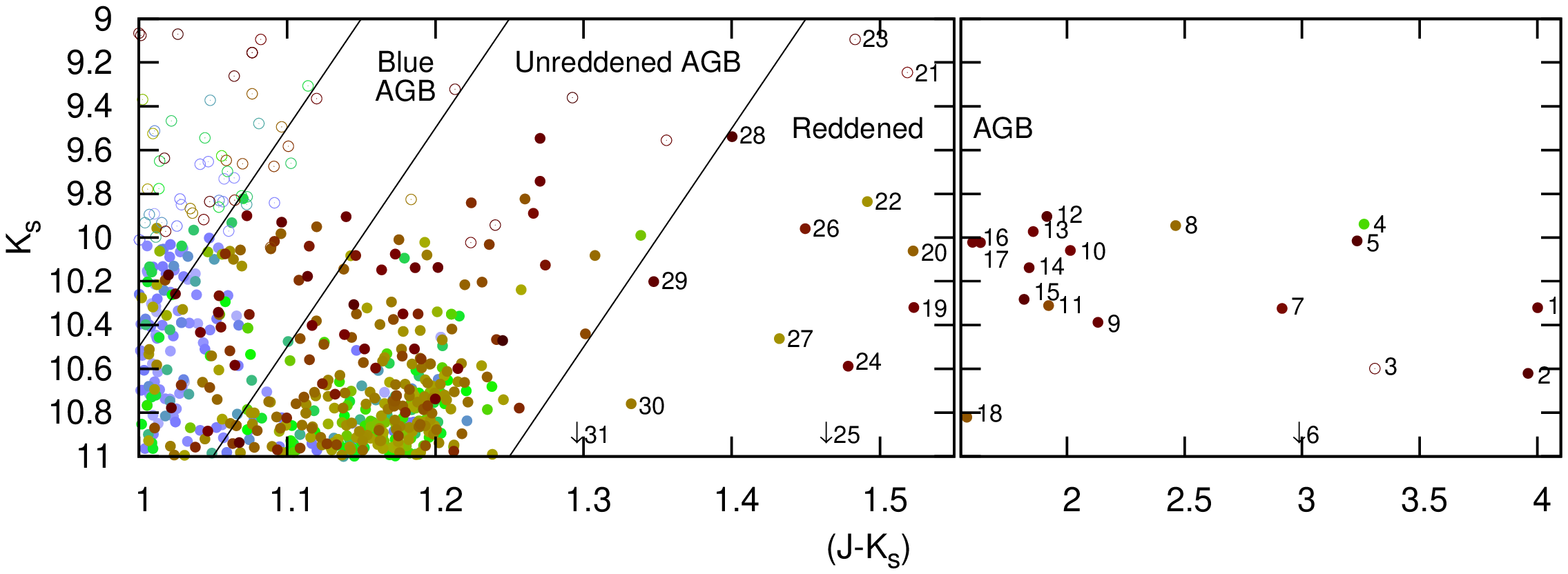}}
\caption{VISTA colour--magnitude diagrams of the Sgr dSph. The colour scale shows the variability in $Z$. The top and left panels show binned averages, including only stars appearing in eight or more epochs which are below the saturation limit. In the top panel, the size of the points corresponds to the number of stars contained within that bin. Boxes show regions expanded in the lower panels. The central panels show close-ups of the Sgr dSph giant branch and its tip. The diagonal lines in the right-hand and lower panels identify the ranges used for separating the blue, unreddened and reddened Sgr dSph giants. The bottom panel shows the reddened stars at the top of the AGB, where the reddest, numbered stars are referred to in the text and have been investigated individually. Open symbols in the right and bottom panels indicate objects close to the saturation limit.}
\label{CMDFig}
\end{figure*}

\begin{figure*}
\centerline{\includegraphics[height=0.95\textwidth,angle=-90]{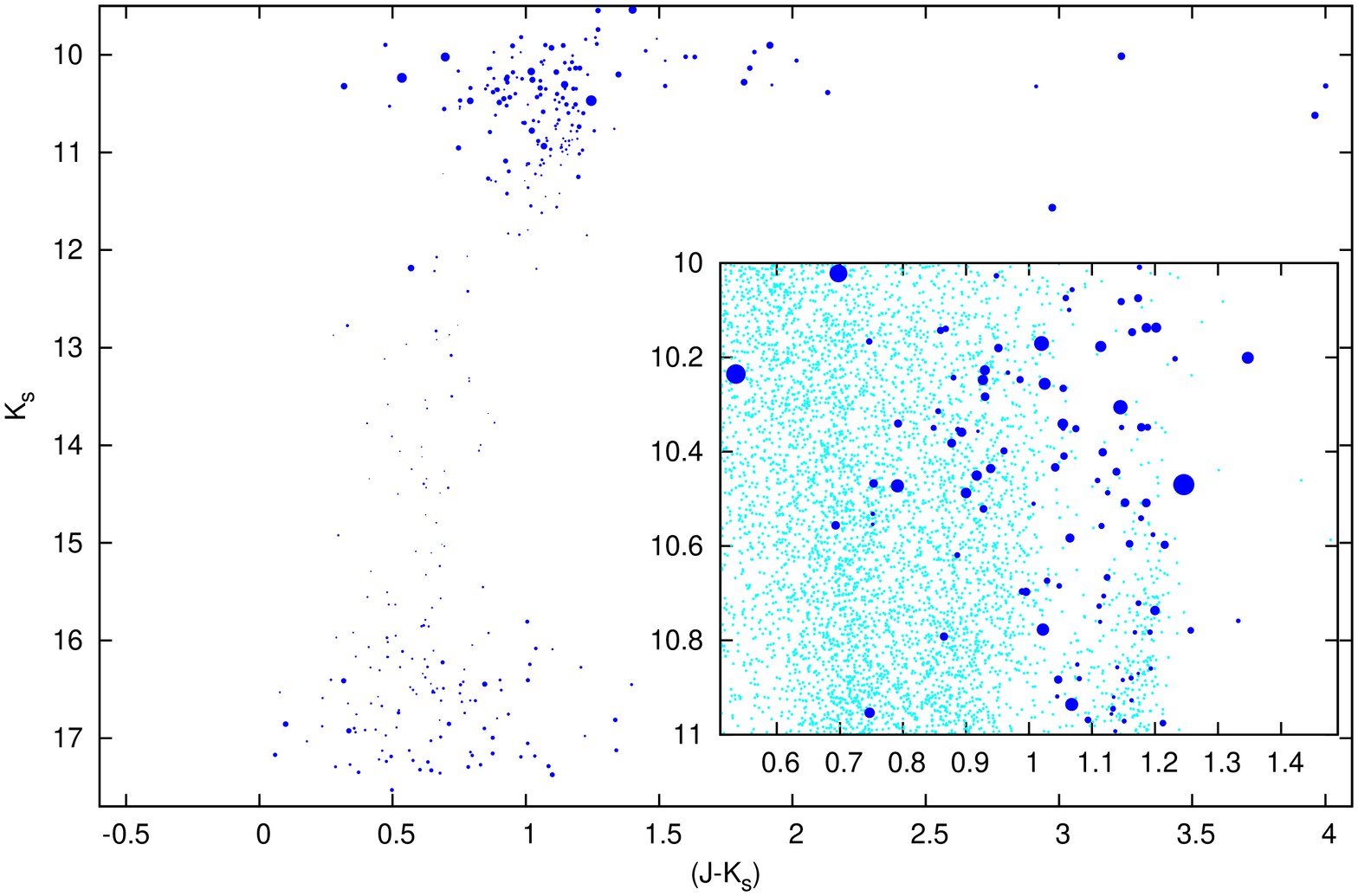}}
\vspace{-5mm}
\centerline{\includegraphics[height=0.95\textwidth,angle=-90]{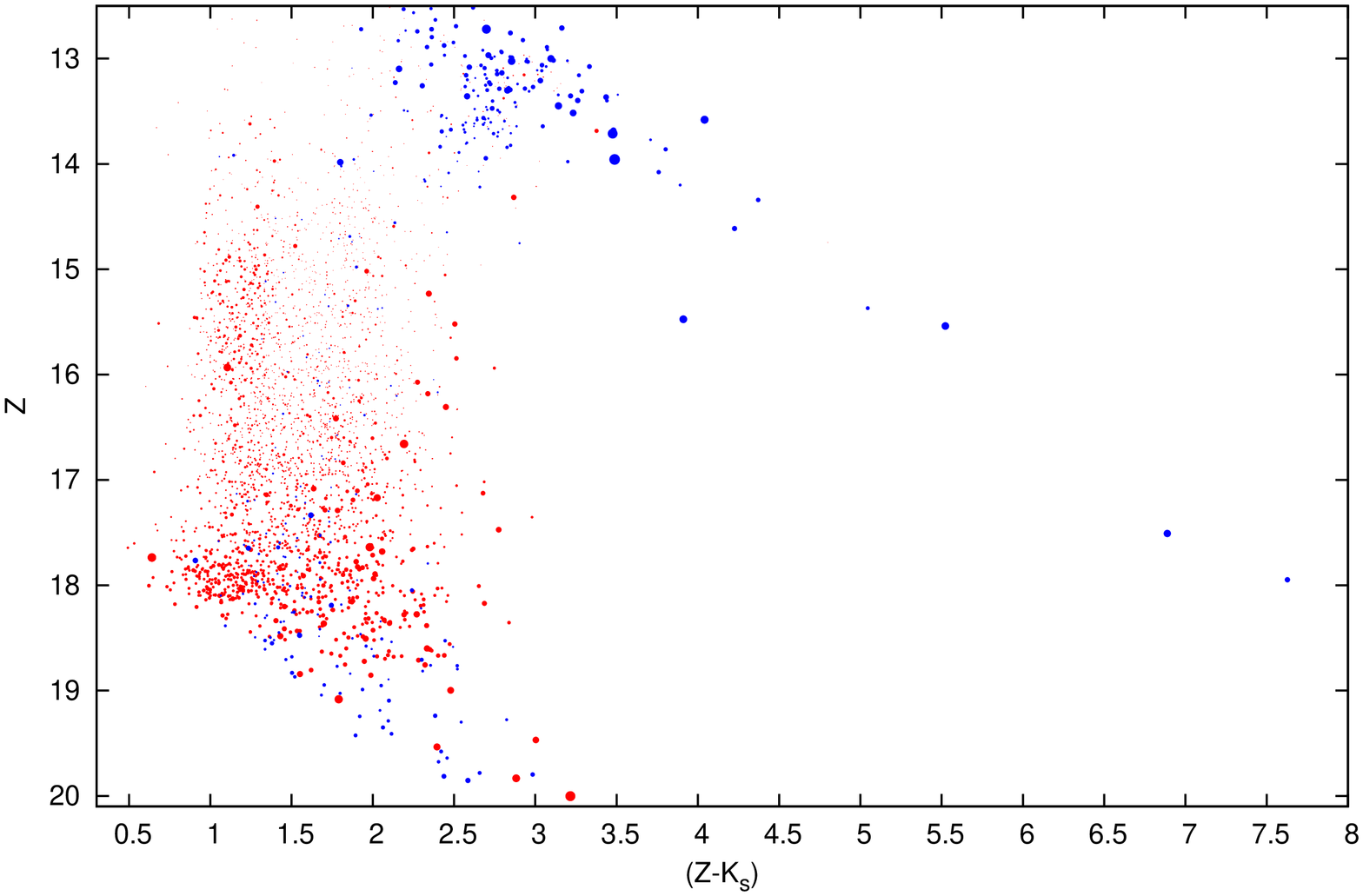}}
\caption{VISTA colour--magnitude diagrams of the Sgr dSph, showing only the variable stars, where symbol size is scaled to the amplitude of variability. Red points denote variables selected on the basis of the standard deviation of their magnitudes (bottom panel only), while blue points denote longer-period variables selected on the basis of comparison of the standard deviations to the point-to-point deviations of their magnitudes. The inset in the top panel shows the region near the tip of the RGB of the Sgr dSph, where stars not identified as variables are shown as cyan points. Stars with mostly saturated data are not plotted.}
\label{VarsFig}
\end{figure*}

Band and tile merging for the $J$- and $K_{\rm s}$-band data were carried out in Paper II, producing a single set of photometry for each source. In Paper II, emphasis was placed on creating the most complete colour--magnitude diagram, but this is not optimised for detecting variability. We re-evaluate the band and tile merging here to place better emphasis on a clean selection of variable stars and extraction of sources that were not detected contemporaneously with the simultaneous $J$- and $K_{\rm s}$-band data.

We begin by concatenating every $Z$-band detection, retaining its epoch, tile number, magnitude and error. Within this dataset, we iteratively merge points into the same object if they fall within a certain radius. This radius increases with each iteration, being initially set at 0$\farcs$125, then 0$\farcs$25 for the second iteration, 0$\farcs$375 for the third, 0$\farcs$5 for the fourth, and 0$\farcs$6 for subsequent iterations. This was found to be necessary to avoid grouping together random points (e.g. from poorly resolved objects or cosmic rays) with points from nearby sources. The largest matching radius is somewhat smaller than the 1$\farcs$2 used in Paper II to merge the three-colour ($ZJK_{\rm s}$) catalogue. We note in that paper that altering the value between 0$\farcs$5 and 2$\farcs$0 has little effect on the final $ZJK_{\rm s}$ catalogue. However, matches at larger radii tend to be spurious, unrelated detections, either not showing a stellar point-spread function or being related to nearby bright or saturated sources. A large radius artificially increases the variability of these sources quite dramatically, thus we seek to reduce this effect. Typically around 6--7 iterations were needed for this process. The average magnitude of each object and standard deviation of detections from the mean of their parent object were then computed. By this method, 34\,517\,215 $Z$-band detections were agglomerated into 4\,267\,464 objects.

We then followed an iterative procedure to remove zero-point variations between tiles and between epochs. The average offset between the detection magnitudes and the parent object magnitudes was computed for each epoch--tile combination. The zero point was adjusted to remove these offsets. Objects were included in this calculation if their parent magnitude was between $13.5 < Z < 17.5$ mag, the standard deviation of their detections was $<$0.4 mag, and the detection was $<$1 mag from the parent object's mean. This process was carried out iteratively, with the last criterion being tightened by 40\% on each iteration (thus $<$0.6 mag, $<$0.36 mag, etc.). Five iterations were needed to reduce the relative differences in the photometric zero point to under 0.001 mag.

Objects were then merged with the $J$ and $K_{\rm s}$ data from Paper II to produce a new three-colour photometric catalogue. Matches were deemed valid if the radius between objects in $Z$ and in $J$+$K_{\rm s}$ was $<$0$\farcs$6. Some 1\,981\,347 $Z$-band objects were matched with $J$- or $K_{\rm s}$-band detections, of which 1\,138\,100 were matched with both.

The baseline of our coverage for individual objects is at least 60 days for most sources, typically 90--100 days, and reaching a maximum of 117 days (Figure \ref{TimebaseFig}). The cadence of our observations are of order 10 days, and we cannot expect to reliably detect periodic variability on shorter timescales. This window of 10--100 days is longer than RR Lyrae or classical Cepheid pulsators, but short compared to the expected pulsation periods of high-amplitude long-period variables (LPVs), which cover many months to years (e.g.\ \citealt{Groenewegen04}). Consequently, we do not expect to obtain meaningful periods for most stars, and our data are expected to underestimate the full range in amplitude of many variables.

\subsection{Removing bad data}
\label{BadDataSect}

As with any large catalogue of data, the resulting dataset contained significant quantities of bad data which necessitated the removal of outliers. For each object, we calculated the median magnitude, and the median deviation of detections from that magnitude. Any detections lying more than 8.5 median deviations from the median were removed. For a normal distribution, this should occur roughly once every 10$^8$ detections, therefore we expect all these outliers to be bad data. This simple cut removed most of the bad data from our sample. Objects appearing in only one epoch were also removed, resulting in a fully reduced catalogue of 3\,777\,091 objects.

The south-eastern detector on VISTA suffers from poor data quality. Where stars fall on this detector in one tile and another detector in a different tile, they acquire an artificially high variability, which we wish to mitigate. For sources lying on multiple tiles, we compute the standard deviation of magnitudes for each tile. If the standard deviation of any one tile's detections is less than half of that of all detections, the tile with the lowest standard deviation (provided it has more than one data point) is taken. Sources with eight or fewer detections were removed from the catalogue, leading to a final catalogue of 2\,669\,072 objects.

Bright objects in the catalogue also suffer from saturation in the VISTA data. Figure \ref{MagVarFig} shows the standard deviation of magnitudes for an object ($\sigma Z$) as a function of the $Z$-band magnitude of that object. The increase that can be seen at brighter magnitudes is due to saturated sources, and a hard cutoff of $Z>12.5$ mag would remove these 7\,400 sources, though they are retained for later investigation. The VISTA integration time was set to achieve maximum signal without saturation at the AGB tip\footnote{The point where temperature begins to increase and the star enters the post-AGB phase.} of the Sgr dSph. With one or two exceptions, the AGB stars in VISTA do not seem to be affected by saturation.

The remaining objects show typically very good sensitivity to variability (Figure \ref{MagVarFig}). The median variability of objects between $12.5 \leq Z \leq 15$ mag is 0.0090 mag, which approximates the scatter in our observations of stars of this magnitude (corresponding to the majority of the upper giant branches).

\subsection{Identifying and classifying variable sources}
\label{GetVarSect}

\begin{table*}
 \centering
 \begin{minipage}{160mm}
  \caption{List of candidate short-period variables in our data, showing the r.m.s.\ variability in the lightcurve ($\delta Z$) and total observed range $<Z>$).}
\label{SPVTable}
  \begin{tabular}{ccccccccccc}
  \hline\hline
   \multicolumn{1}{c}{ID}	&    \multicolumn{1}{c}{RA}	&    \multicolumn{1}{c}{Dec}
& \multicolumn{1}{c}{$Z$}	& \multicolumn{1}{c}{$\delta Z$}	& \multicolumn{1}{c}{$<Z>$}	
& \multicolumn{1}{c}{$J$}	& \multicolumn{1}{c}{$K_{\rm s}$}\\
\ & \multicolumn{1}{c}{(deg)} &\multicolumn{1}{c}{(deg)} &
\multicolumn{1}{c}{(mag)} & \multicolumn{1}{c}{(mag)} & \multicolumn{1}{c}{(mag)} & \multicolumn{1}{c}{(mag)} & \multicolumn{1}{c}{(mag)} & \ \\
 \hline
SPVSgr18462138--2927051 & 281.5891 & --29.4514 & 10.569 & 0.585 & 2.243 & 9.785 & 9.130 \\
\nodata & \nodata & \nodata & \nodata & \nodata & \nodata & \nodata & \nodata \\
\hline
\end{tabular}
\end{minipage}
\end{table*}

\begin{table*}
 \centering
 \begin{minipage}{160mm}
  \caption{List of bright long-period variables in our data, with symbols as in Table \ref{SPVTable}.}
\label{LPVTable}
  \begin{tabular}{ccccccccccc}
  \hline\hline
   \multicolumn{1}{c}{ID}	&    \multicolumn{1}{c}{RA}	&    \multicolumn{1}{c}{Dec}
& \multicolumn{1}{c}{$Z$}	& \multicolumn{1}{c}{$\delta Z$}	& \multicolumn{1}{c}{$<Z>$}	
& \multicolumn{1}{c}{$J$}	& \multicolumn{1}{c}{$K_{\rm s}$}\\
\ & \multicolumn{1}{c}{(deg)} &\multicolumn{1}{c}{(deg)} &
\multicolumn{1}{c}{(mag)} & \multicolumn{1}{c}{(mag)} & \multicolumn{1}{c}{(mag)} & \multicolumn{1}{c}{(mag)} & \multicolumn{1}{c}{(mag)} & \ \\
 \hline
LPVSgr18403594--3125445 & 280.1498 & --31.4290 & 13.673 & 0.031 & 13.412 & 12.179 & 11.112 \\
\nodata & \nodata & \nodata & \nodata & \nodata & \nodata & \nodata & \nodata \\
\hline
\end{tabular}
\end{minipage}
\end{table*}

\begin{table*}
 \centering
 \begin{minipage}{160mm}
  \caption{List of candidate faint long-period variables in our data, with symbols as in Table \ref{SPVTable}.}
\label{FLPVTable}
  \begin{tabular}{ccccccccccc}
  \hline\hline
   \multicolumn{1}{c}{ID}	&    \multicolumn{1}{c}{RA}	&    \multicolumn{1}{c}{Dec}
& \multicolumn{1}{c}{$Z$}	& \multicolumn{1}{c}{$\delta Z$}	& \multicolumn{1}{c}{$<Z>$}	
& \multicolumn{1}{c}{$J$}	& \multicolumn{1}{c}{$K_{\rm s}$}\\
\ & \multicolumn{1}{c}{(deg)} &\multicolumn{1}{c}{(deg)} &
\multicolumn{1}{c}{(mag)} & \multicolumn{1}{c}{(mag)} & \multicolumn{1}{c}{(mag)} & \multicolumn{1}{c}{(mag)} & \multicolumn{1}{c}{(mag)} & \ \\
 \hline
FLPVSgr18402211--3055111 & 280.0922 & --30.9198 & 17.913 & 0.020 & 17.208 & 16.977 & 16.645 \\
\nodata & \nodata & \nodata & \nodata & \nodata & \nodata & \nodata & \nodata \\
\hline
\end{tabular}
\end{minipage}
\end{table*}

An automated classification was made to identify individual variable soures in the final catalogue. While the approach we use is ad hoc, it is generally effective at identifying objects of interest, as it naturally biases our results to finding the LPVs on the Sgr dSph giant branches.

We begin by splitting the magnitude--variability diagram (Figure \ref{MagVarFig}) into 0.1 mag bins. For each bin, compute a median variability and standard deviation about the median. As the distribution of variability within each bin does not follow a normal distribution, we take the standard deviation to be difference between the mean variability and the point where 84.15 per cent of objects in that bin have lower variability. Based on the number of objects within that bin, we compute the number of standard deviations from the median where we would expect the furthest outlier to lie: e.g.\ 2 st.\ dev.\ for 44 objects, 3 st.\ dev. for 741 objects, 4 st.\ dev.\ for 31\,574 objects, etc. Any object lying beyond this boundary (the dashed line in Figure \ref{MagVarFig}) is classified as a candidate variable.

Some 3\,056 objects meet this criterion, of which 30 are brighter than our $Z = 12.5$ mag saturation cutoff. We consider these to be candidate variables of periods both long and short, compared to our observing cadence (typical time between observations). However, with only $\sim$12 epochs, it is not possible to conclusively identify variability in variables of periods close to or shorter than our observing cadence, as their light curves are indistinguishable from noise. While this catalogue is biased towards bright objects, it counter-intuitively is dominated by much fainter objects, due to their overwhelming numbers. We list these objects in Table \ref{SPVTable}.

For this reason, we derive a second list of longer-period variables: those with periods greater than our observing cadence. Such variables will have magnitudes that follow trends with time (whatever those trends may be), so will vary less between adjacent epochs than they do over the whole dataset. We determine this long-period variability for each object by comparing the standard deviation of magnitudes over all epochs to the average magnitude difference between adjacent epochs (i.e.\ $(|m_2-m_1| + |m_3-m_2| + \cdots + |m_n-m_{n-1}|)/(n-1)$, where $m_i$ is the $Z$-band magnitude at epoch $i$ and $n$ is the number of epochs). For noise-dominated variability, the ratio of the standard deviation to the average epoch-to-epoch variation should be 0.857 on average, and is unlikely to reach below 0.51 for a set of data of our size. We therefore consider any object with a ratio of $<$0.52 to be a candidate LPV.

There are 674 objects matching this criterion. Of these, we only include the 337 objects where both $J$- and $K_{\rm s}$-band data are also available. These objects are mostly dominated by the bright, red giants of magnitude $K_{\rm s} < 12$ mag (176 objects), but include a large number of fainter objects with $17.6 < K_{\rm s} < 16$ mag (121 objects). Such faint stars are not likely to show these kind of long-period variations: they are typically lower-RGB stars in the Sgr dSph or sub-giant stars in the Galactic Bulge, and are too small in radius to oscillate at long periods. We do not consider it likely that they are truly variable. However, we have not been able to identify any instrumental reason why they should appear variable. As a result, we have split this list of objects into a bright and faint list, with an arbitrary cutoff at $K_{\rm s} = 12.5$ mag. We list the bright variables in Table \ref{LPVTable} and the faint candidate variables in Table \ref{FLPVTable}.

\section{Results}
\label{ResultsSect}

\subsection{Overview}
\label{OverviewSect}

Figure \ref{CMDFig} presents a binned, colour-encoded colour--magnitude diagram of the final catalogue. Figure \ref{VarsFig} plots the identified variables individually on the colour--magnitude diagram. 

The minimum of the average variability among bins of a given magnitude is a rough proxy for the noise associated with stars of that magnitude. The background noise in the data increases with magnitude, as the signal-to-noise ratio decreases. It is also typically less on the well-populated branches than it is in the less-populated regions of the diagram. This is commensurate with objects of poor-quality photometry being scattered from the main features of the diagram in all three bands. Such poor-quality photometry is most often due to contamination by nearby bright stars.

Variability increases towards the top of the Sgr dSph giant branch (the top of the Bulge giant branch is saturated). The central panels of Figure \ref{CMDFig} show the variability on the Sgr dSph giant branch increasing above the background noise around $K_{\rm s} = 12.5$ mag, and continuing to increase to $\sim$0.02--0.03 mag at the RGB tip, where virtually every star is identified as a variable. AGB stars brighter than the RGB tip show larger variability, as expected for their higher luminosity. Variability is particularly prevalent in the extended, reddened sequence of carbon stars (see Paper I) that are shown in the bottom panel of Figure \ref{CMDFig}. The observed variabilities for the AGB stars are less than can be expected for such stars due to the relatively short period of observations (Section \ref{CalcVarSect}).

We have divided the AGB population up into three components, based on colour--magnitude cuts, which are given in Figure \ref{CMDFig}. We showed in Paper I that, typically, stars on the red side of the AGB are carbon stars, stars in the middle of the AGB are oxygen-rich stars, and stars on the blue side of the AGB are metal-poor oxygen-rich stars. However, stars on the blue side of the AGB could be Bulge giants, and variability means that stars may migrate through different regions during their pulsation cycle. Differential reddening across the field may also shift sources slightly, but given the observed extinction ($E(B-V) = 0.15$ mag for M54; \citealt{Harris96}) and its smoothness across the \emph{COBE} maps \citep{SFD98}, this is likely to have very minor effects at $J$- or $K$-band, where extinction is 3--8 times less \citep{Gray92}.

\subsection{The carbon-rich reddened AGB}
\label{RAGBSect}

Paper I showed that the extended sequence of stars lying redward of the AGB tip are all likely to be carbon-rich. We have shown these 31 stars in the bottom panel of Figure \ref{CMDFig}, sorted numerically by ($J-K_{\rm s}$) colour and limited by the criterion $(J-K_{\rm s}) > (23.5 - K_{\rm s})/10$, used in Paper I. These sources are also listed in Table \ref{RAGBTable}, where the observed variability, $\sigma Z$, should be viewed as a lower limit due to the comparatively short baseline of coverage.

We investigate these sources individually in the Appendix, but note that six are not automatically identified by the criteria for variability presented in Section \ref{GetVarSect}. These are RAGB 3, 21 and 23 (saturated stars which may be incorrectly placed on the colour--magnitude diagram), RAGB 4 (where we probably only observe a short fraction of the period around maximum light), RAGB 18 (which has poor photometry due to its position on both the tile edge and the malfunctioning VISTA chip) and RAGB 27 (a genuine low-amplitude variable). Thus, of the 26 sources where we can reliably detect variability, 25 are variable. Of these, 21 are identified among the LPVs (Table \ref{LPVTable}) and 21 among the general variables (Table \ref{SPVTable}).

The reddened AGB sources, previously shown to be mostly or wholly carbon-rich (Paper I) appear to be dominated by large amplitude variables with periods much greater than the $\sim$100-day coverage period. \citet{BD13} published reliable near-infrared periods and amplitudes for reddened stars in the Sgr dSph, but they only cover three of the reddened AGB stars in our field of view (RAGB 7, 15 and 29 in the Appendix). For RAGB 7 and 15, it appears that we have observed most of the light curve's range at $Z$-band. For RAGB 29, we appear to only cover around half the expected amplitude.

\subsection{The unreddened AGB}
\label{UAGBSect}

\begin{figure}
\centerline{\includegraphics[height=0.47\textwidth,angle=-90]{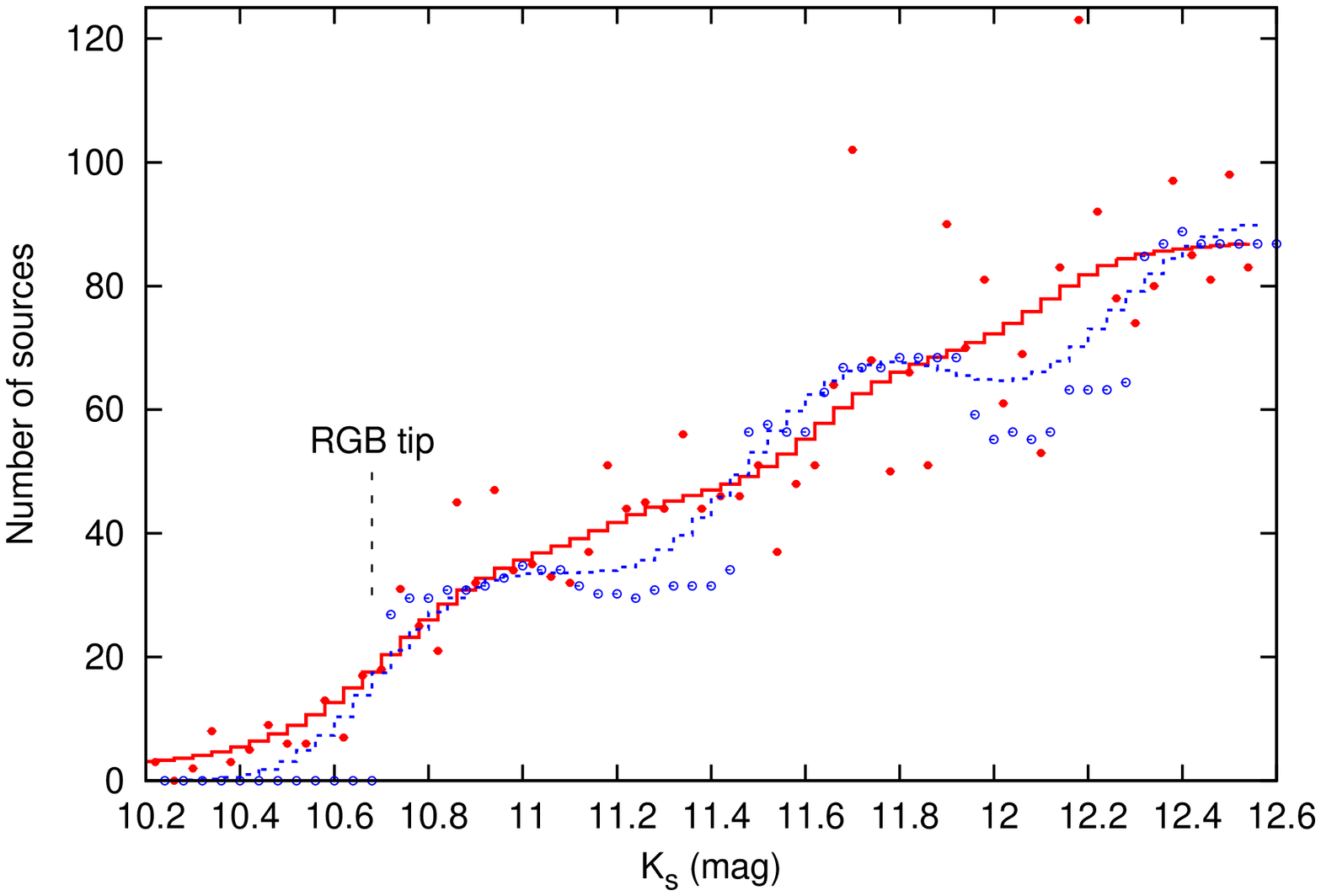}}
\centerline{\includegraphics[height=0.47\textwidth,angle=-90]{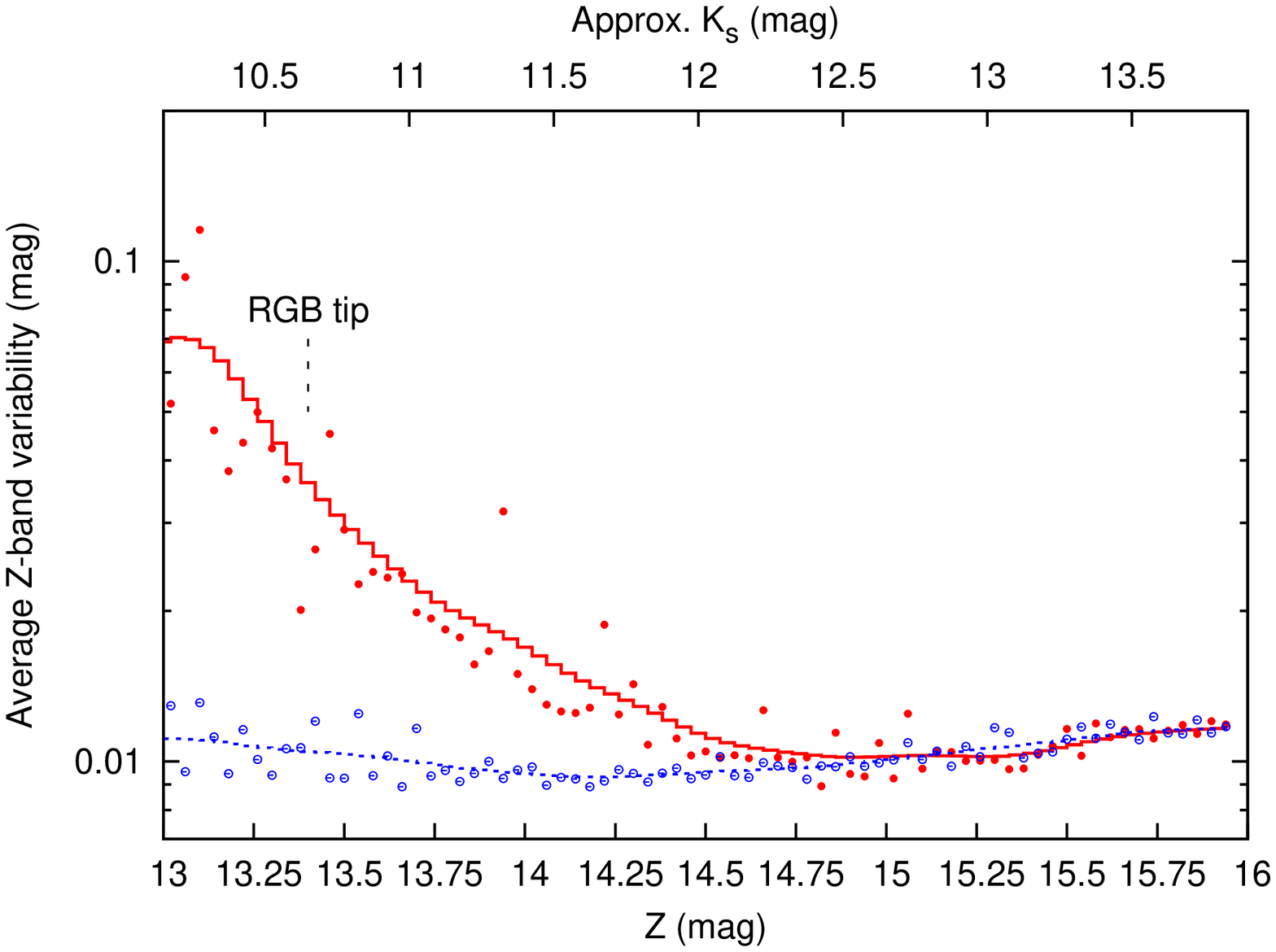}}
\caption{Top panel: a histogram of all colour-selected sources around the RGB tip of Sgr dSph, regardless of their variability (red filled points), and a comparative a Padova model luminosity function (blue hollow points). The solid, red line and the dashed, blue line give (respectively) both datsets smoothed by a 0.1 mag Gaussian function. Bottom panel: $Z$-band variability of those sources (red filled points). The comparative variability from the Bulge population, which is dominated by noise, is shown as the blue, hollow points. The lines show the datasets similarly smoothed by 0.1 mag. The approximate corresponding $K_{\rm s}$-band magnitude, valid for the Sgr dSph stars only, is shown on the upper axis. Spikes in both datasets are due to single highly variable sources among small numbers of stars in those bins.}
\label{RGBVarHistFig}
\end{figure}

Stars in the Sgr dSph above the RGB tip ($K_{\rm s} = 10.68$ mag; for spectral typing information, see Paper I) are greatly dominated by oxygen-rich AGB stars. Stars in this region can be as evolved as the carbon-rich AGB stars described above, but do not show the same reddened appearance as their silicate dust does not absorb strongly in $J$ and $K_{\rm s}$ bands (see Paper I, figure 10). We showed in Paper I that most Sgr dSph giants should evolve into carbon stars, but that does not preclude some very-evolved oxygen-rich objects lying in this region.

The RGB tip in the Sgr dSph is visually obvious from our colour--magnitude diagram (Figure \ref{CMDFig}), as it is in the 2MASS colour--magnitude diagram of the wider galaxy (Paper I). However, the relatively small number of stars near the RGB tip and their intrinsic variability make defining its precise magnitude difficult. We are hampered by the fairly small number of stars near the RGB tip, as well as small variations (a few hundredths of a magnitude) in colour and magnitude that arise from slight changes in metallicity and extinction across the galaxy (see Paper II). Despite being the brightest stars, the most variable AGB stars also have sufficient intrinsic variability to cross the RGB tip during the pulsation cycle. In keeping with Papers I and II, we have chosen to place it at $K_{\rm s} = 10.68$ mag, where the source density per magnitude bin roughly halves. A histogram of sources with magnitude (Figure \ref{RGBVarHistFig}) suggests sources typically scatter around the RGB tip by $\approx$0.10 mag (the shape of the RGB tip being only slightly changed when a 0.10 mag smoothing is applied). Thus, we can be reasonably confident that sources above $K_{\rm s} = 10.55$ mag are AGB stars in the Sgr dSph (at $\gtrsim$68 per cent confidence, accounting for the fraction of Sgr dSph non-members in that region).

Also shown in Figure \ref{RGBVarHistFig} is an 8-Gyr, $Z = 0.004$ luminosity function from the Padova models (from \citet{MGB+08} with the \citet{GWG+10} Case A formulism for AGB mass loss). The luminosity function is poorly reproduced in the model, due to the noise in these fast-evolving stages of the underlying evolutionary model, but the slope of the RGB is qualitatively reproduced, indicating our photometric data are largely complete. Indeed, we expect the completeness limit to typically lie near $K_{\rm s} = 15.5$ to 16.5 mag, depending on the crowding in that region.

Defining a blue edge to the upper AGB is also not trivial. A large number of stars scatter blueward of the giant branches, and presumably represent either the metal-poor AGB stars within the Sgr dSph galaxy, or variable AGB stars from the bulk population at the blue end of their pulsation cycle. At the lowest metallicities around [Fe/H] = --1.7 dex, the giant branches become diluted among the redder giants in the Galactic Bulge (Paper II).

Variability is significantly enhanced on the upper AGB. Variability, as derived from the standard deviation of photometry for each source, is typically $\sigma Z$ = 0.072 mag. This compares to 0.011 mag for sources at the same $Z$-band magnitude in the Bulge, and 0.019 mag for sources one $Z$-band magnitude fainter on the RGB. With sufficient data to obtain a period, this enhanced variability would offer a method of extracting the metal-poor AGB populations from the confusion of the Galactic Bulge, though we cannot accurately do this with our present data. Figure \ref{RGBVarHistFig} shows a histogram of the variability on the upper RGB and AGB, where we note that the average $Z$-band variability of stars brighter than $Z \approx 14.5$ mag is well-fit by a power law. Stars fainter than this limit are likely variable, but are below the level of the noise in our data. We discuss this power law further in Section \ref{VarAmpSgrSect}.

There are 62 sources brighter than $Z = 14.5$ mag which match the colour limits for giant branch stars in the Sgr dSph ($(21.5 - K_{\rm s})/10 < (J-K_{\rm s}) < (23.5 - K_{\rm s})/10$) applied in Paper I. We list these in Table \ref{UAGBTable}, and examine the more-interesting sources among them in the Appendix. Of these 62 objects, 34 were identified as candidate variables by our criteria, and 24 as longer-period variables. Five stars were saturated at $Z$, and one is a known non-member. Upon visual investigation, the light curves of the remaining stars typically show some longer-period variability, indicating that the automated routines do not flag every variable source. This appears to be because the light curve modulations in these objects are very small, where the scatter in individual points is comparable to the light curve amplitude.

Detailed investigation of each star reveal that the stars in the unreddened AGB have shorter periods than those on the reddened AGB (see the notes in the Appendix, including Table \ref{UAGBTable}). \citet{BD13} find periods of between 210 and 396 days for some of the reddest carbon stars in the Sgr dSph. \citet{WMIF99} report periods of 228--360 days. Conversely, variations in the lightcurves of the unreddened AGB stars seem to contain many stars with periods of $\sim$100 days, with some perhaps as short as 40 days. The period--luminosity relation for LPVs (e.g.\ \citealt{MW85,KB03,ITM+02,ITM+04}) therefore suggests that unreddened (mostly oxygen-rich) stars either tend to be less luminous, and/or are overtone pulsators.

Pulsations in the unreddened stars are also weaker, with the average $\sigma Z$ being 0.075 mag for unreddened stars (excluding the probable non-members and saturated stars) and 0.103 mag for reddened stars. As we preferentially appear not to cover the entire range of variability for the reddened stars, the real difference is likely to be considerably larger.

\subsection{The blue AGB}
\label{BAGBSect}

Figure \ref{CMDFig} identifies the stars classified as the blue AGB, which lie between the Sgr dSph and Bulge giant branches. This region contains a substantially higher fraction of variable stars compared to Bulge stars of similar magnitude, but a similar fraction compared to the Sgr dSph. We would therefore expect that the majority of objects falling into this gap are variable, blue AGB stars from the metal-poor populations of the Sgr dSph.

Any LPVs from the Bulge or foreground populations should either have periods much shorter than the observing cadence (typically between one and a few days in the case of single, pulsating stars) or lie on the period--luminosity sequence E. The origins of sequence E are unclear, but it is generally suggested to arise from a contact binary population \citep{OW03,WOK04,DKB+06,Soszynski07,MSO+08,NWCS09,MvLDB10}.

The RGB tip in these metal-poor populations is expected to be considerably fainter in $K_{\rm s}$ magnitude than the bulk population (see, e.g., Paper II). We assume a value here of $K_{\rm s} = 10.8$ mag. Of the 139 stars classified as blue AGB stars (Figure \ref{CMDFig}), 36 show variability of $\sigma Z > 0.024$ mag (cf.\ the background level of around 0.008 mag), of which 28 are classified as variables. Table \ref{BAGBTable} lists these 28 variables, and four other variables which were identified visually, but were missed by our automatic variable finder.

Examination of these objects shows no obvious indications that they are metal-poor or not. Table \ref{BAGBTable} indicates a mixture of carbon- and oxygen-rich stars. Examination of the 2MASS colours of individual objects shows that many are scattered here due to their variability: variability at $J$-band is often considerably greater than at $K_{\rm s}$-band (see, e.g., \citealt{BD13}), hence at peak $J$-band output the stars can have abnormally blue $(J-K_{\rm s})$ colours.

Figure \ref{CMDFig} shows a general absence of blue AGB stars at magnitudes fainter than $K_{\rm s} = 11.4$ mag, with most of the stars here seeming to come from the bluer Bulge stars. We therefore suspect that most of the objects we list as blue AGB stars are simply normal AGB stars scattered here by their variability, rather than metal-poor members of the Sgr dSph. We may expect some of these objects to be Type II Cepheid variables in the post-AGB phase of evolution, though we cannot identify these without periods and better-sampled lightcurves.

\subsection{The red giant branch tip}
\label{SARGSect}

Stars in the few magnitudes below the RGB tip will contain a mixture of RGB and AGB stars. These stars are not heavily affected by blending and are easily bright enough that we should detect every object. The similarity of the surface properties of RGB and AGB stars near the RGB tip means we cannot distinguish between the two branches at these magnitudes, and should therefore obtain unbiased comparisons of RGB versus AGB properties.

Of the 1\,548 stars on the giant branch between $10.55 < K_{\rm s} < 12$ mag, 68 are identified as variable, 47 of which are longer-period variables. However, the larger r.m.s.\ scatter in this region compared to Bulge stars of similar brightness (Figure \ref{CMDFig}) suggests that these are only the more-variable end of a continuous distribution that extends to far lower amplitudes. Indeed, the average stellar variability is still detectable down to $K_{\rm s} \approx 12$ at an r.m.s.\ variability of $\delta Z \gtrsim 0.01$ mag (Figure \ref{RGBVarHistFig}). We discuss the amplitude of variability in this magnitude range in greater detail in Section \ref{VarAmpSgrSect}.

Variables near the RGB tip of the Large Magellic Cloud tend to occupy {\rm the first and second harmonic sequences (Wood's sequences B and A)} on the period--luminosity diagram, with periods between roughly 10 and 60 days (see, e.g., \citealt{Wood00,ITM+04,SUK+04,TSI13}). Stars preferentially shift to the higher harmonic, with periods of 10--20 days, with the majority of the stars lying on sequence A below $K_{\rm s} \approx 13$ mag. Assuming stars in the Sgr dSph follow the same trend, this transition should occur {\rm by} $K_{\rm s} \approx 11.5$ mag in the Sgr dSph. Our observing cadence is such that the {\rm period of stars on sequence A become shorter than twice our observing cadence at a similar magnitude. Such stars will therefore not have structured lightcurves in our data, and will} not be identified as longer-period variables in our analysis. This could explain the lack of longer-period variables fainter than $K_{\rm s} \approx 11.5$ mag ($Z \approx 14$ mag; Figure \ref{VarsFig}).

\subsection{Variability on the horizontal branch}
\label{HBSect}

\begin{figure}
\centerline{\includegraphics[height=0.47\textwidth,angle=-90]{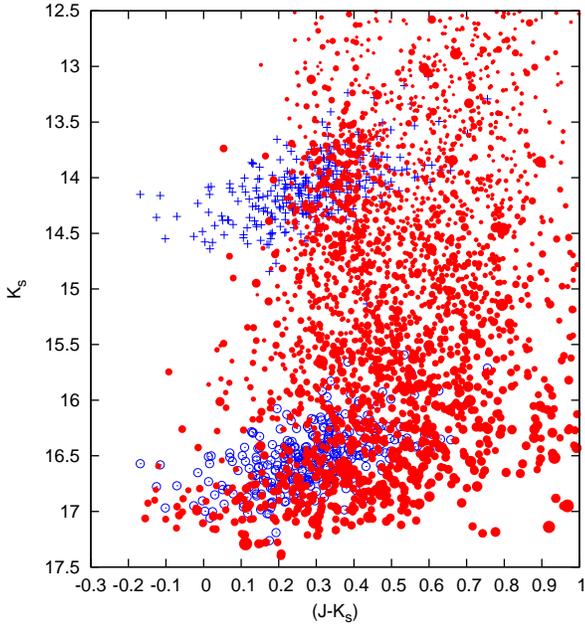}}
\caption{Variable stars identified in the Sgr dSph VISTA data are shown as red dots, with symbol size proportional to their $Z$-band variability. For comparison, RR Lyrae from globular clusters are shown as blue, open circles, which have been shifted to match the distance and extinction of the Sgr dSph; and as blue + signs as scaled to the distance of the Galactic Bulge.}
\label{RGBVarCMD2}
\end{figure}

The horizontal branch of the Sgr dSph is expected to contain a significant number of RR Lyr stars. The $Z$--$J$ and $Z$--$K_{\rm s}$ colours of RR Lyrae stars are yet to be investigated in any great detail. However, RR Lyr stars are warm enough that $Z$ covers the Rayleigh--Jeans tail and the spectrum should be free of molecular opacity, thus the $JHK_{\rm s}$-band variation should serve as a reasonable proxy. \citet{SCA+08} compute variabilty of $\delta (JHK) \approx 0.25$ mag for RR Lyr itself, thus this is the approximate sensitivity limit we must reach to detect RR Lyrae. Figure \ref{MagVarFig} shows that our short-period variable catalogue should contain objects this variable down to $Z \approx 19$ mag.

It is well known that only stars from ancient populations lose sufficient mass to become hot enough to fall in the instability strip on the horizontal branch (e.g.\ \citealt{DCJ+08}). Even in globular clusters, stars of similar metallicity to the bulk population of the Sgr dSph ([Fe/H] = --0.7 to --0.4 dex; \citealt{SDM+07}) do not become hot enough to become RR Lyrae (e.g.\ \citealt{CMD+01,MBvL+11}) and it is not until metallicities comparable to the Sgr dSph's metal-poor population ([Fe/H] $\approx$ --1.2 dex; \citealt{SDM+07}) that RR Lyrae begin to appear (e.g.\ \citealt{CMD+01,BMvL+09,MvLD+09}).

We have selected 2MASS objects labelled as RR Lyrae stars in clusters from the SIMBAD astronomical database. There are 338 objects with $JHK_{\rm s}$ photometry, of which a small fraction appear to be misclassified variables of other types and several have properties which do not place them in the clusters with which they are associated. We have corrected the 2MASS $J$ and $K_{\rm s}$ magnitudes for distance and reddening to the values of the Sgr dSph (taken as 25 kpc, with $E(B-V)$ = 0.15, $E(J-K_{\rm s})$ = 0.025 and $A_Ks$ = 0.04 mag) and plotted them in Figure \ref{RGBVarCMD2}. The RR Lyrae location clearly overlaps with a significant number of variables identified in the Sgr dSph. The region $0.10 < (J-K_{\rm s} < 0.55$ and $15.95 < K_{\rm s} < 16.95$ contains $\approx$75\% of the globular cluster RR Lyrae \citep{CMD+01} and 324 of our candidate variable stars. While not all these will be RR Lyrae (many of them will be Galactic Disc eclipsing binaries), a significant proportion of them should be, as should a number of stars for which we do not have $J$- and $K_{\rm s}$-band data, but do have $Z$-band lightcurves. Better sampling of stars' lightcurves and deeper $JK_{\rm s}$-band observations are needed to identify properties for these stars.

\subsection{Variability in the Bulge and foreground}
\label{BulgeSect}

Our observations should also easily be sensitive enough to detect RR Lyrae in the Galactic Bulge. The Bulge red clump itself lies near $(Z-K_{\rm s}) \approx 1.3$ mag, $Z \approx 14.0$ mag, where our sensitivity limit is $\delta Z \approx 0.01$ mag. However, the observed red clump represents the amalgam of the hydrogen-burning RGB bump and the helium-burning red clump. The stars of interest, the hotter counterparts of the red clump stars, are likely to be bluer. Taking a distance modulus of $\Delta Z = 2.42$ mag between the Sgr dSph and the Bulge, we derive the region containing RR Lyrae to cover the range $0.10 < (J-K_{\rm s} < 0.55$ and $13.53 < K_{\rm s} < 14.53$. 

The above region of our colour--magnitude diagram (Figure \ref{RGBVarCMD2}) contains 340 candidate variable stars. Many of these have higher amplitudes than stars in neighbouring areas of the colour--magnitude diagram and are thus likely Bulge RR Lyrae stars. We note an absence of the bluest RR Lyrae stars, which correspond to the lowest-mass stars (more accurately, the lowest masses of stellar envelopes). Thus, either the oldest Bulge population we sample is not as old as the globular cluster population or, perhaps more likely, the enhanced evolutionary rate caused by globular cluster self-pollution of helium does not apply in the Bulge \citep{CDOY+13,CCD+13}.

The foreground population lies at varying distances from us, thus we do not expect any significant clusters of foreground variables in the colour--magnitude diagram. Indeed, Figure \ref{VarsFig} shows candidate variables on the blue edge of the populated colour--magnitude diagram. A possible enhancement of short-period variability is seen between 1.0 $\lesssim (Z-K_{\rm s}) \lesssim$ 1.5 mag and 15 $\lesssim K_{\rm s} \lesssim$ 16 mag. Without further data it is not possible to determine the nature or veracity of these candidate variables with accuracy.

\begin{figure}
\centerline{\includegraphics[height=0.47\textwidth,angle=-90]{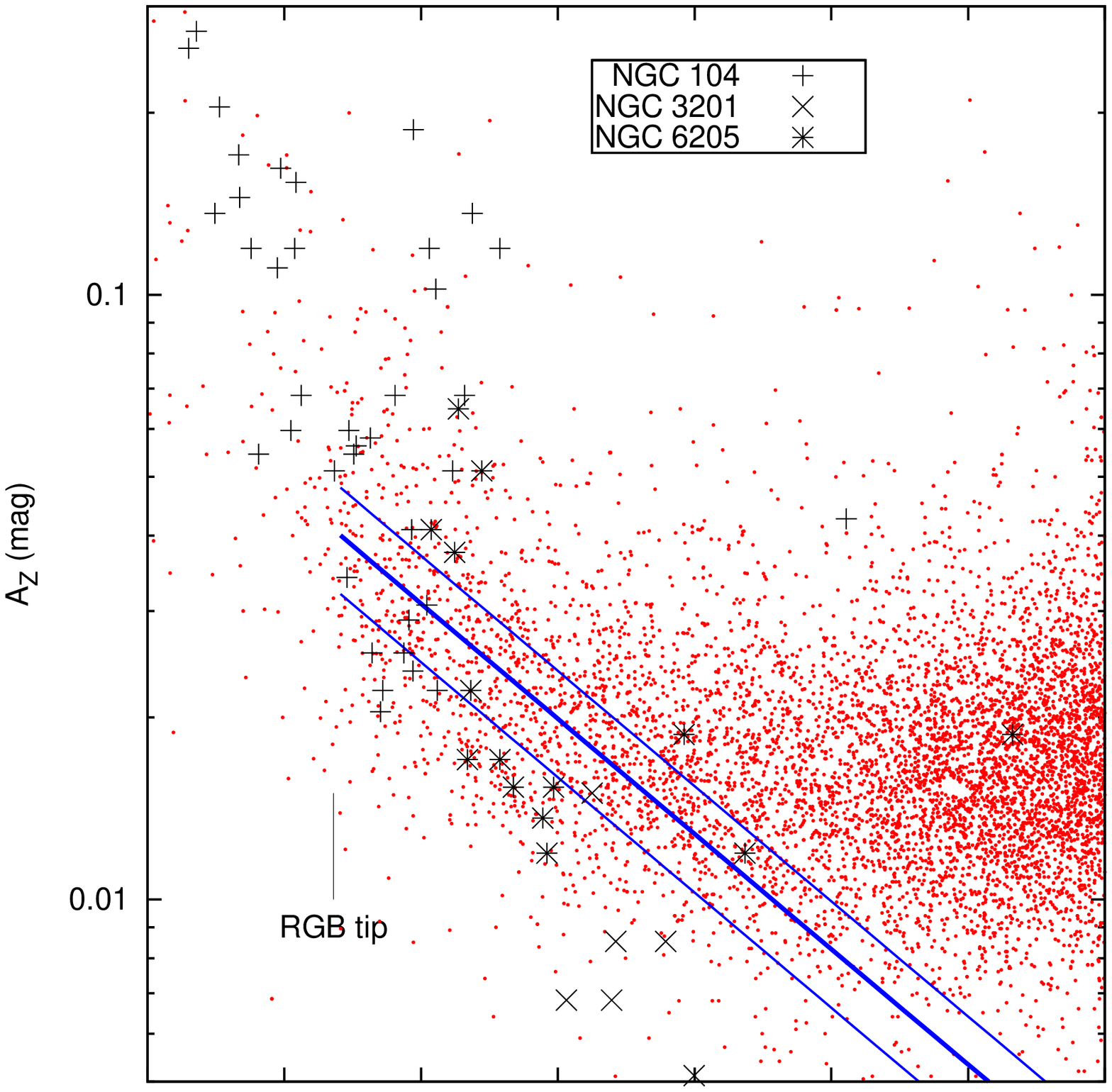}}
\centerline{\includegraphics[height=0.47\textwidth,angle=-90]{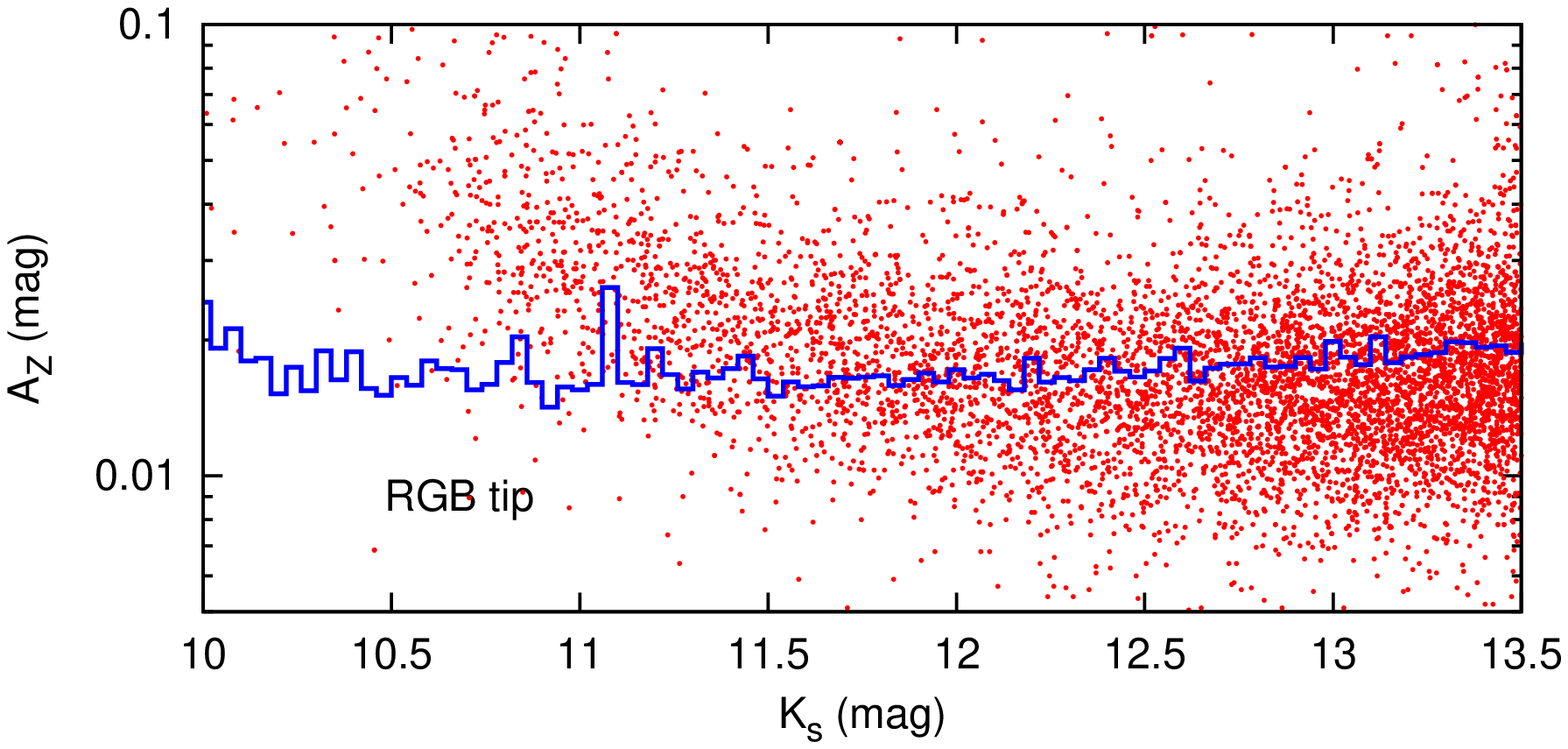}}
\caption{Semi-amplitudes of colour-selected stars on the Sgr dSph giant branches (red dots). Top panel: The blue lines show our scaling of the \protect\citet{KB95} relation to our data with an approximate error margin. Black points show the semi-amplitudes at $V$ (scaled appropriately in both axes) for stars in globular clusters (see figure legend). Bottom panel: the blue, dotted line shows the approximate observational noise, represented by a histogram of amplitudes for Bulge stars, shifted by 0.5 mag to account for their difference in ($Z-K_{\rm s}$) colour.}
\label{RGBVarHistFig2}
\end{figure}

\subsection{Amplitude of variability among Sgr dSph giants}
\label{VarAmpSgrSect}

Our observations cover a variety of giant branch variables, from Mira-like pulsators to variables at the limit of detection. Figure \ref{RGBVarHistFig2} plots the variability of every star selected by colour to lie on the Sgr dSph giant branches. Near the RGB tip, virtually every star is observed to be variable above the noise. At fainter magnitudes, detection of variability appears limited by stochastic noise in the observations. The lack of a noticeable difference in amplitude as one crosses the giant branch tip suggests that the amplitude of these variables depends only on the temperature and/or luminosity of the star, and not on whether the object is an RGB or AGB star, or on which pulsation sequence the star lies. This is in agreement with the theory and observations of solar-like pulsations, which we discuss below.

Objects with these comparatively minor variations in brightness are semi-formally classified as small amplitude red giants, or SARGs. While SARGs have been known for some years (e.g.\ \citealt{Grenon93,JMSM97}), the event of precision variability surveys such as \emph{Hipparcos} (\citealt{vLEG+97}; updated \citealt{vanLeeuwen07}), MACHO \citep{Wood00} and OGLE \citep{SUK+04} have led to their discovery in abundance. \citet{TSI13} estimate masses of SARGs in the Large Magellanic Cloud as being $\sim$0.9--1.4 M$_\odot$. A lack of well-sampled photometry prevents us from making a similar estimate, though this mass range is consistent with the mass expected for a Sgr-dSph-like population over a wide range of ages ($\sim$2.6--12 Gyr; \citealt{MGB+08,GWG+10}).

The origin of the variability in red giant branch stars is poorly understood. It likely comes from a combination of stochastically excited solar-like oscillations driven by convection, which produce true radial changes in the star (e.g.\ \citealt{BBB+11}), and $\kappa$-mechanism-driven opacity changes near the hydrogen ionisation boundary \citep{Cox80,OC86,UOA+89}, which can then cause proportionally much larger changes in visual light output when stars become cool enough to form temperature-sensitive, optically absorbing molecules (e.g. TiO, C$_2$; \citealt{RG02}).

For both processes, we expect the amplitude of variability to increase at shorter wavelengths (due to both the blackbody flux changes and the generally stronger molecular bands at bluer wavelengths), while variability on the Rayleigh--Jeans tail ($\lambda \gtrsim 1.5$ $\mu$m) should be roughly constant. Indeed, this is roughly what is observed in multi-wavelength studies (e.g.\ \citealt{MvLDB10,PSK+10,BD13}). However, the frequency of the pulsations is not well-defined, and is generally described in Fourier space as a central peak with a Lorentzian envelope \citep{KSB06}.

The lack of a single, clearly defined frequency makes it difficult to compare the semi-amplitudes we find (where we may only have one or two periods) with amplitudes from the literature, which are usually derived from sums over longer-term power spectra (e.g.\ \citealt{KB95,DS10}). \citet{KB95}, however, note the proportionality of the pulsation amplitude (in magnitudes) to stellar fundamental parameters (their equation 8):
\begin{equation}
A_\lambda \propto \frac{L}{\lambda T_{\rm eff}^2 M} ,
\label{AEq1}
\end{equation}
where $A_\lambda$ is the amplitude at wavelength $\lambda$, and $L$, $T_{\rm eff}$ and $M$ are the stellar luminosity, effective temperature and mass, respectively.

We fit this relation to our $Z$-band amplitudes in Figure \ref{RGBVarHistFig2}, assuming the $K_{\rm s}-L$ and $K_{\rm s}-M$ relations of an appropriate Dartmouth isochrone (\citealt{DCJ+08}, assuming [Fe/H] = --0.55, [$\alpha$/Fe] = --0.2, $t$ = 6 Gyr; \citealt{SDM+07}; Paper II). The r.m.s.\ scatter of $\sim$20\% around this fit is expected to be attributable to the limited period of our observations and errors in our determination of the stellar luminosity and effective temperature. Although we have biased our sample by making a colour-selected subset of the Sgr dSph giants (Figure \ref{CMDFig}), this selection is wide enough to cover the bulk populations of the Sgr dSph, which have the potential to span quite a range in age and metallicity (Paper II). The narrowness of the amplitude--magnitude relation suggests an r.m.s.\ scatter in stellar mass for the stars on the upper RGB of the Sgr dSph bulk population of $\ll$20\%.

\section{Discussion}
\label{DiscSect}

\subsection{Carbon richness and variability}
\label{SpecVarSect}

\begin{figure}
\centerline{\includegraphics[height=0.47\textwidth,angle=-90]{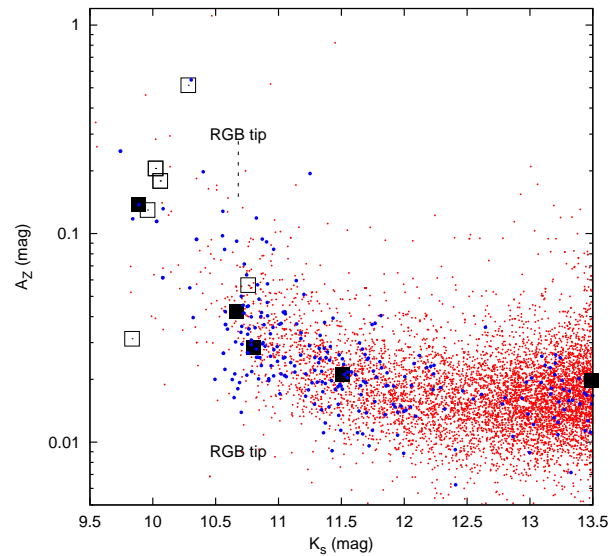}}
\caption{As Figure \ref{RGBVarHistFig2}, where objects have been highlighted by surface chemistry (Paper I). Smaller, blue circles mark M-type stars while larger, black squares make carbon stars. Stars have been colour selected to lie on the Sgr dSph giant branches, though we include stars redder than the Sgr dSph giant branch (i.e.\ probable carbon stars) as hollow squares.}
\label{RGBVarHistFig3}
\end{figure}

\begin{figure}
\centerline{\includegraphics[height=0.47\textwidth,angle=-90]{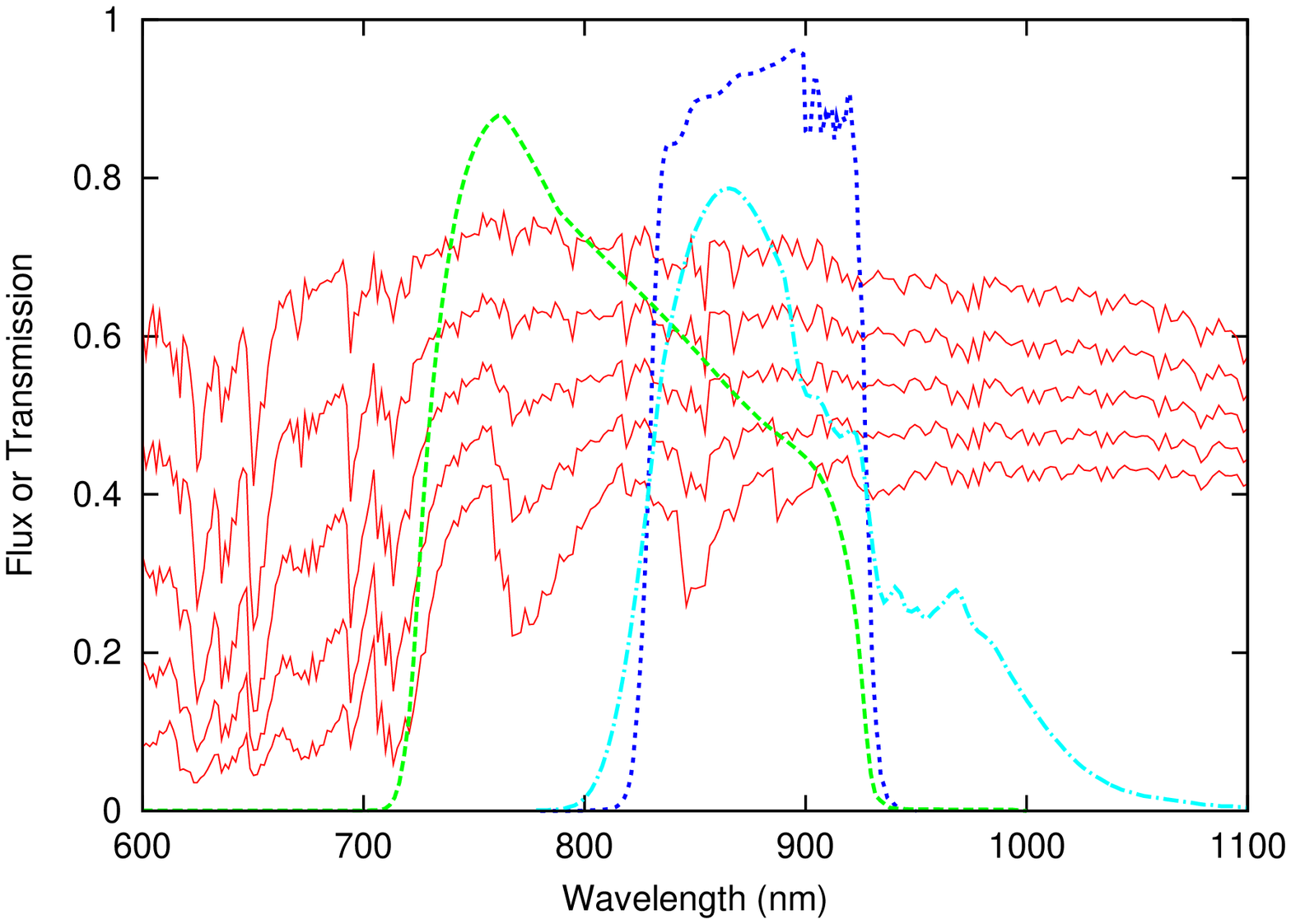}}
\centerline{\includegraphics[height=0.47\textwidth,angle=-90]{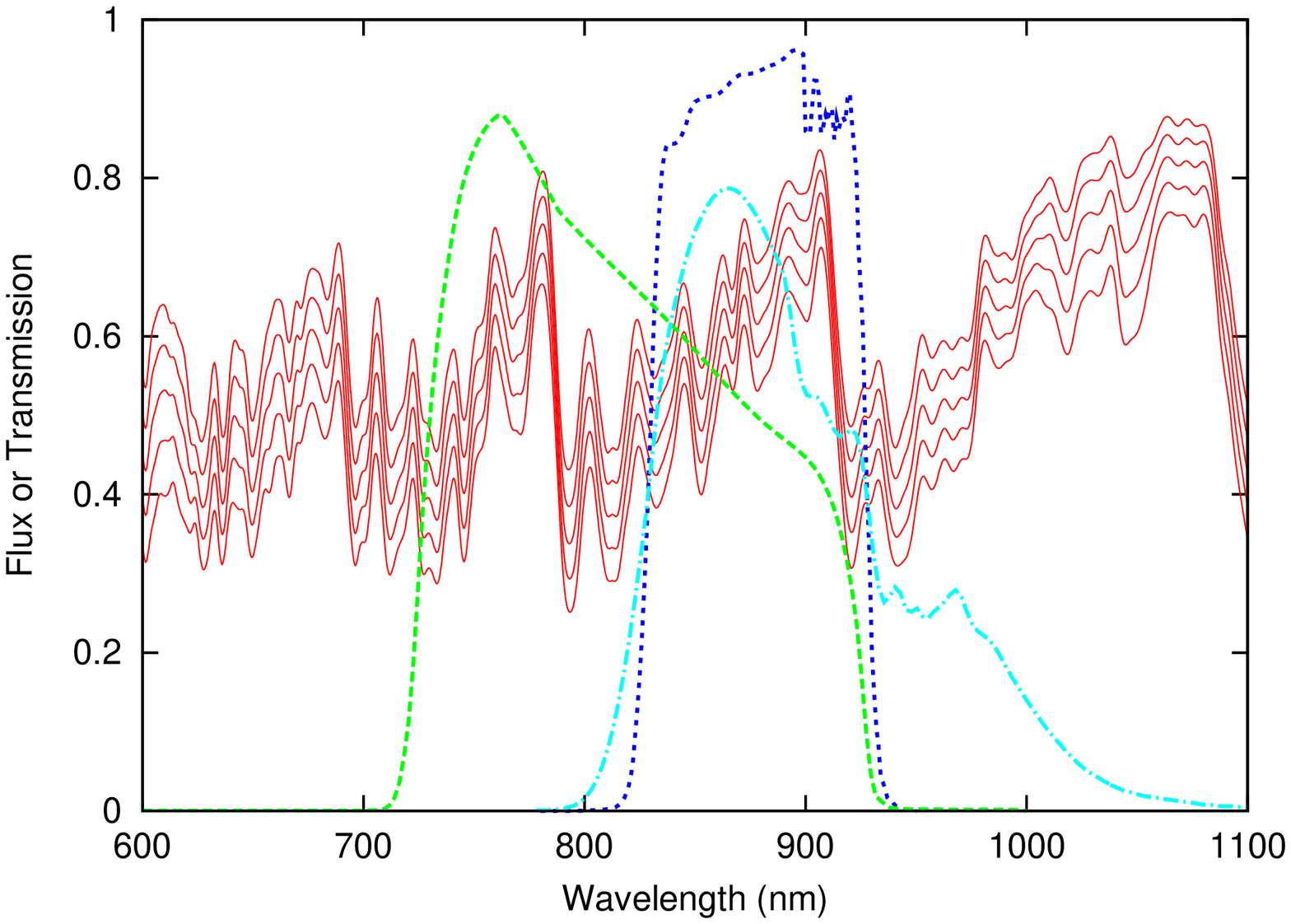}}
\caption{Top panel: oxygen-rich \protect{\sc BT-Settl} model atmospheres \protect\citep{AGL+03} for [Fe/H] = --0.5 dex, log($g$) = 0.5 dex stars for $T$ = 3400 K (bottom, solid red lines) to 3800 K (top) in steps of 100 K. The blue dashed line shows the VISTA $Z$-band filter, while the other lines show the Cousins $I$-band filter (left, long-dashed green) and Sloan $z$-band filter (right, dot-dashed cyan) for comparison. Bottom panel: as top panel, for {\sc comarcs} carbon-rich models \protect\citep{AGN+09} for [Fe/H] = --0.48 dex, log($g$) = 0 dex, C/O = 1.4, $M$ = 2 M$_\odot$.}
\label{ZFig}
\end{figure}

Figure \ref{RGBVarHistFig3} shows the amplitudes of carbon- and oxygen-rich giants, as derived from optical spectra in Paper I. While the number of carbon stars in the sample is small, they are generally seen to follow the same trends as the oxygen-rich stars. The $Z$ band is relatively free of molecular opacity in both oxygen- and carbon-rich stars compared to other optical filters (Figure \ref{ZFig}). A weak TiO band exists at 8432\AA, while a moderately weak CN band is present at 9148 \AA, the latter lying only partially within the $Z$ filter transmission. The $Z$ band therefore is probably the bluest useful broadband filter which can trace effective temperature changes among both oxygen- and carbon-rich stars.

The lack of difference between oxygen- and carbon-rich star variability suggests that changes in molecular absorption do not contribute significantly to the observed variability. This similarity also extends to the brightest objects above the RGB tip. These stars have temperatures of $\approx$3500 K, at which point molecular effects have become significant in other parts of the spectrum, and where substantial variability ($A_{\rm Z} \approx 0.1$ mag) exists among both oxygen- and carbon-rich AGB stars.

While the carbon stars identified through optical spectra show no obvious difference in variability, our spectral survey (Paper I) preferentially selected those stars which were likely to be optically bright: i.e.\ those which have little circumstellar extinction. Optically faint stars, with considerable extinction from circumstellar dust, may show different properties, which we investigate in the following section.

\subsection{Dust production and variability}
\label{DustVarSect}

\begin{figure}
\centerline{\includegraphics[height=0.47\textwidth,angle=-90]{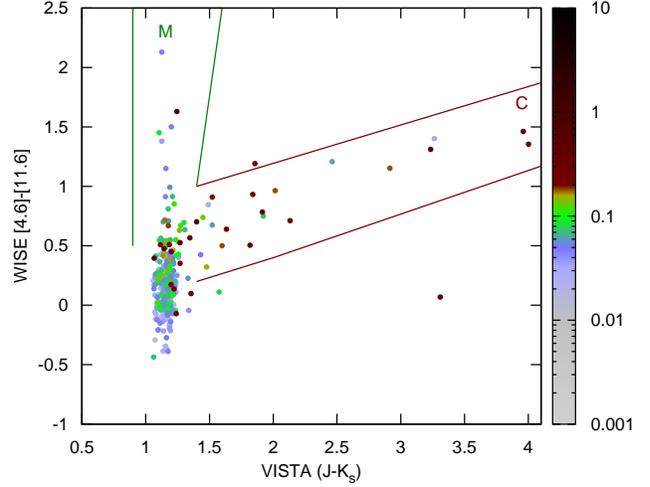}}
\caption{Colour--colour diagram showing VISTA ($J-K_{\rm s}$) (high values are indicative of carbon-rich dust) versus \emph{WISE} [4.6]--[11.6] colours (high values are indicative of oxygen-rich silicate dust). The colour scale denotes the semi-amplitude of variability at $Z$ in magnitudes.}
\label{DustVarFig}
\end{figure}

\begin{figure}
\centerline{\includegraphics[height=0.47\textwidth,angle=-90]{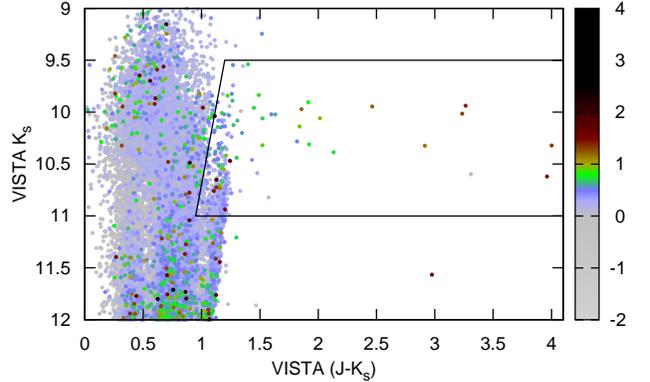}}
\caption{Colour--magnitude diagram showing VISTA near-infrared data, coloured by \emph{WISE} [4.6]--[11.6] mid-infrared colour, showing stars with dust excess. The {\rm solid} line shows the stars used in Figure \protect\ref{DustVarFig}.}
\label{DustVar2Fig}
\end{figure}

Figure \ref{DustVarFig} shows a colour--magnitude diagram showing our derived ($J-K_{\rm s}$) colour versus the [4.6]--[11.6] colour of \emph{WISE}, reproducing figure 10 of Paper I but incorporating the $Z$-band variability data. Only the giant branch stars above $K_{\rm s} = 11.5$ mag are shown. Carbon stars, producing carbon-rich dust, scatter towards large ($J-K_{\rm s}$) colours. Silicate-producing, oxygen-rich stars scatter towards large [4.6]--[11.6] colours. We remind the reader that we do not cover the entire pulsation period for stars more luminous than the RGB tip. We have previously shown that dust production typically starts around 0.7 mag below the RGB tip, becoming common by the RGB tip \citep{MBvL+11,MvLS+11,MZB12}. We thus expect to underestimate the amplitudes for dust-producing stars, which will show considerable scatter. When averaged across several stars, however, there should still be a correlation between observed amplitude and the actual variability over the full pulsation period.

While Figure \ref{RGBVarHistFig3} shows that there is no significant difference between unreddened carbon and oxygen-rich stars, Figure \ref{DustVarFig} shows that dusty carbon stars with near-infrared excess are significantly more variable than dusty oxygen-rich stars with mid-infrared excess. While the difference between carbon and oxygen-rich stars is surprising (we return to this point below), it is the comparative lack of variability in the oxygen-rich stars with large \emph{WISE} [4.6]--[11.6] colours that is most unexpected.

We argue that these stars are artefacts or oddities, and do not typically represent the galaxy's highly evolved oxygen-rich stars. Figure \ref{DustVar2Fig} shows that \emph{WISE} [4.6]--[11.6] colours of 0.5--1.0 mag are typical among AGB stars above the RGB tip (most of which are oxygen-rich; see Paper I). However, within the marked area, stars unreddened at ($J$--$K_{\rm s}$) with \emph{WISE} [4.6]--[11.6] $>$ 1 mag typically fall below the RGB tip. While these stars may be undergoing a unique phase of evolution (e.g.\ envelope stripping by a companion or a thermal pulse), such stars occur sporadically across the colour--magnitude diagram, and their \emph{WISE} [11.6]--[22] colours are slightly larger than would be expected for dusty stars (typically they are 1.1--1.7 mag, compared to the expected $\lesssim$1.0 mag). This suggests these objects are artefacts: they may either have bad photometry in the lower-resolution \emph{WISE} catalogue or may be normal stars blended with background infrared-bright galaxies (cf.\ \citealt{BMvL+08,MvLD+09}). A visual inspection of the most extreme objects shows no evidence of a resolved source in the optical or infrared, nor a positional offset between 2MASS and \emph{WISE}.

On this basis, the oxygen-rich stars in the Sgr dSph may reach a maximum of \emph{WISE} [4.6]--[11.6] $\approx$ 1.0 mag in the Sgr dSph, after which they either reach the AGB tip or become carbon stars. We therefore cannot draw conclusions between pulsation input to stellar winds for highly evolved oxygen-rich stars in general. In the Sgr dSph, the relative lack of oxygen-rich Miras (stars with $\delta Z \gtrsim 1$ mag) hints that stars evolve into carbon stars before they can become luminous enough to become Mira pulsators. We caution, however, that we do not cover the entire amplitude range of many of the Miras we would expect. The exception to this may be the star 2MASS 18423574--3018390, which is among the most variable stars in our sample. Despite lying just above the RGB tip in our VISTA data, it essentially defines the AGB tip in the 2MASS data, implying an photometric range of over a magnitude at $K_{\rm s}$ band.

The lack of enhanced variability among oxygen-rich stars with [4.6]--[11.6] $\gtrsim$ 0.6 suggests that dust production does not correlate with variability, hence pulsations do not contribute significantly to the driving of winds from oxygen-rich stars in the Sgr dSph. In the carbon-rich stars, however, pulsation amplitude and ($J$-$K_{\rm s}$) colour are correlated, suggesting that pulsations may provide significant input to driving the stellar wind.

\subsection{Clues on the $\kappa$ mechanism}
\label{DustVarSect}

The similarity of $Z$-band amplitudes among RGB and AGB stars (Figure \ref{RGBVarHistFig2}) and carbon and oxygen-rich stars (Figure \ref{RGBVarHistFig3}) implies that the driving mechanism of the pulsation does not strongly depend on either the internal structure of the star, nor the molecules formed at its surface. In the $\kappa$-mechanism model (e.g.\ \citealt{OW05}), the surface pulsations are primarily excited stochastically by convective cells, causing the periodic motion of a hydrogen ionisation front below the surface of the star. All modes are excited, with the emerging frequencies controlled by the level of damping or enhancement those harmonics experience. Typically, damping decreases as stars become cooler and larger, hence the observed pulsation amplitude increases (\citealt{HO03}, Section 2.6).

At some point, the atmosphere will become sufficiently extended such that molecules and eventually dust can form in the outer layers. Two effects can then occur to increase the observed optical pulsation. Firstly, the changing temperature of the star during the pulsation cycle changes the equilibrium of molecule and dust formation. This causes a correlated, periodic change in opacity from the change in column density of molecules and dust in the extended atmosphere, leading to an amplification of the pulsation. This is termed the \emph{external} $\kappa$ mechanism. The resonant timescale for this is typically several times longer than the internal $\kappa$ mechanism, possibly leading to longer, secondary pulsations (e.g.\ \citealt{WFGS94,HFD95}), but it is likely to impart extra amplitude to the primary pulsation period. Secondly, the increased molecular opacity cools the star, bringing optical bands around $Z$ closer to the Wien tail of the star's spectral energy distribution. Identical temperature changes in a cooler star would lead to greater amplitudes being recorded.

It is difficult to determine whether cooling of the star or increased molecular opacity has the greater effect. Only the enshrouded carbon stars do not follow the power-law increase of variability with $K_{\rm s}$-band magnitude shown in Figure \ref{RGBVarHistFig2}. The enshrouded stars do not appear on Figure \ref{RGBVarHistFig3} as they were too faint in the optical to be observed in the spectral catalogue of Paper I. In these obscured stars, $K_{\rm s}$-band magnitude is no longer a reasonable proxy for luminosity, {\rm as they are cool enough and obscured enough that the peak of the spectral energy distribution approaches the $K_{\rm s}$-band. They therefore have a higher bolometric luminosity} than oxygen-rich stars at the same $K_{\rm s}$ magnitude, but are also more luminous than any star that is likely to be oxygen-rich, with the exception of 2MASS 18423574--3018390. Furthermore, the amplitudes of stars at or above the RGB tip may be best described as lower limits, as our observing period is unlikely to cover their entire amplitudes. However, there is a lack of noticable steps in amplitude as stars become cool enough to form molecules or dust. This implies that the external $\kappa$ mechanism can likely only be effective in stars where the dust layers are optically thick: the dust-enshrouded carbon stars. It may be this external $\kappa$ mechanism that allows the pulsations to become more effective and help drive the wind in carbon stars.

\section{Conclusions}
\label{ConcSect}

We have reported the near-infrared variability of over a million stars towards the Sgr dSph, covering $\approx$11 square degrees of the sky with VISTA. We have concentrated our efforts on resolving long-period variability at the galaxy's giant branch tip, where we have obtained an r.m.s.\ sensitivity of $\approx$15 mmag in $Z$. From these data, we obtain the following main conclusions:
\begin{itemize}
\item Variability increases with increasing luminosity as a power law, as is seen in the Magellanic Clouds and Solar Neighbourhood.
\item No difference in variability is seen between stars on the RGB and AGB. This is as expected for stochastic ($\kappa$-mechanism) excitation within the upper stellar atmosphere, as the exciting convective motion is not strongly affected by the deeper structure of the star.
\item Carbon stars unreddened by dust do not show significantly more variability than their oxygen-rich counterparts. Again, this is expected if the mechanism exciting pulsation is, e.g., a moving ionisation front of hydrogen in the atmosphere.
\item Dust-producing oxygen-rich stars do not show higher variability than dustless stars, suggesting that pulsation does not assist their winds. However, the relatively weak infrared excess and low ampltidue of the dusty oxygen-rich stars means we cannot comment on highly evolved stars: stars in the Sgr dSph become carbon rich before this phase.
\item However, the highest variabilities are found in the dust-enshrouded, heavily reddened carbon stars. This may be due to additional varying opacity source (dust) on top of the variation in molecular opacity that occurs during the pulsation cycle, otherwise known as the external $\kappa$ mechanism.
\item Hence, pulsation does not appear to correlate with dust production in oxygen-rich stars, while it does in carbon stars. This suggests that pulsations are only effective in helping to drive the wind from carbon-rich stars in the Sgr dSph.
\end{itemize}

\section*{Acknowledgements}
Based on observations made with ESO Telescopes at the La Silla Paranal Observatory under programme ID 089.D-0113.

The authors thank Prof.\ Tim Bedding for his helpful comments on the manuscript, and the anonymous referee for their helpful comments, which have improved this work.

This publication makes use of data products from the Two Micron All Sky Survey, which is a joint project of the University of Massachusetts and the Infrared Processing and Analysis Center/California Institute of Technology, funded by the National Aeronautics and Space Administration and the National Science Foundation.

The DENIS project has been partly funded by the SCIENCE and the HCM plans of the European Commission under grants CT920791 and CT940627. It is supported by INSU, MEN and CNRS in France, by the State of Baden-W\"urttemberg in Germany, by DGICYT in Spain, by CNR in Italy, by FFwFBWF in Austria, by FAPESP in Brazil, by OTKA grants F-4239 and F-013990 in Hungary, and by the ESO C\&EE grant A-04-046.

This publication makes use of data products from the Wide-field Infrared Survey Explorer, which is a joint project of the University of California, Los Angeles, and the Jet Propulsion Laboratory/California Institute of Technology, funded by the National Aeronautics and Space Administration.


This research has made use of the SIMBAD database, operated at CDS, Strasbourg, France.

\appendix

\section{Notes on individual stars}

\subsection{Reddened AGB stars}


\begin{table*}
 \centering
 \begin{minipage}{160mm}
  \caption{Variability of surveyed red AGB stars.}
\label{RAGBTable}
  \begin{tabular}{rlllllrrcc}
  \hline\hline
   \multicolumn{1}{c}{RAGB}	&    \multicolumn{1}{c}{RA}	&    \multicolumn{1}{c}{Dec}
& \multicolumn{1}{c}{$Z$}	& \multicolumn{1}{c}{$\sigma Z^{(1)}$}	
& \multicolumn{1}{c}{($J-K_{\rm s}$)}	& \multicolumn{1}{c}{$K_{\rm s}$}	
& \multicolumn{1}{c}{$n_{\rm epochs}$}	& \multicolumn{1}{c}{Spectral}
& \multicolumn{1}{c}{Lightcurve}\\
   \multicolumn{1}{c}{ID}	&    \multicolumn{1}{c}{Degrees}&    \multicolumn{1}{c}{Degrees}&
\multicolumn{1}{c}{(mag)} & \multicolumn{1}{c}{(mag)} & \multicolumn{1}{c}{(mag)} & \multicolumn{1}{c}{(mag)} & \multicolumn{1}{c}{\ } & \multicolumn{1}{c}{type$^{2}$} & \multicolumn{1}{c}{shape$^{3}$} \\
 \hline
1	& 285.0089564	& --30.5929859	& 17.948	& 0.1771	& 4.001	& 10.321	& 15	& 	& $\searrow$\\
2	& 285.6323997	& --30.7506886	& 17.509	& 0.3302	& 3.961	& 10.620	& 12	& 	& $\nearrow$\\
3	& 284.7215023	& --31.1800492	& 10.852	& 0.2648	& 3.310	& 10.598	& 18	&n$^4$	& $\rightarrow^5$\\
4	& 284.6378763	& --30.3255223	& 16.463	& 0.0175	& 3.264	&  9.939	& 11	& 	& $\searrow ?$\\
5	& 281.1290016	& --30.6193736	& 15.538	& 0.3564	& 3.234	& 10.015	& 13	&Cm$^b$	& $\nearrow$\\
6	& 285.7912336	& --30.3436439	& 15.475	& 0.3721	& 2.975	& 11.565	& 11	&	& $\nearrow$\\
7	& 283.4207004	& --29.5730037	& 15.369	& 0.0914	& 2.915	& 10.323	& 11	& 	& $\nearrow$\\
8	& 281.2734824	& --29.2659161	& 14.745	& 0.0364	& 2.463	&  9.945	& 12	& 	& $\cup$\\
9	& 285.4702994	& --30.5441979	& 14.612	& 0.1675	& 2.132	& 10.386	& 12	&Cm$^c$	& $\cup$\\
10	& 282.9210818	& --30.0604919	& 13.861	& 0.1068	& 2.016	& 10.059	& 11	&Cm$^{ac}$& $\nearrow$\\
11	& 282.7961712	& --31.3656507	& 14.201	& 0.0501	& 1.923	& 10.310	& 11	& 	& $\cup$\\
12	& 284.9018599	& --30.2626636	& 13.000	& 0.3087	& 1.915	&  9.904	& 12	& 	& $\cap$\\
13	& 281.6345010	& --29.8967807	& 14.342	& 0.1207	& 1.857	&  9.972	& 13	& 	& $\searrow$ or $\cup$\\
14	& 286.2041046	& --31.1816675	& 13.398	& 0.1817	& 1.840	& 10.138	& 12	& 	& $\cap$\\
15	& 283.3724178	& --29.6400384	& 13.515	& 0.2817	& 1.819	& 10.282	& 12	&Cm$^{ac}$& $\nearrow$\\
16	& 283.7191432	& --30.3487050	& 13.309	& 0.1382	& 1.634	& 10.023	& 9	&Cm$^a$	& $\nearrow$ (or $\cap$?)\\
17	& 280.9901262	& --30.6900773	& 13.062	& 0.1283	& 1.600	& 10.021	& 13	& 	& $\cap$\\
18	& 283.4215033	& --31.1300547	& 13.737	& 0.1573	& 1.576	& 10.821	& 31	& 	& $\rightarrow$\\
19	& 286.2695674	& --30.5695651	& 14.078	& 0.1123	& 1.523	& 10.319	& 11	& 	& $\cup$\\
20	& 282.6307102	& --31.0443312	& 13.770	& 0.0401	& 1.522	& 10.060	& 10	& 	& $\searrow$ (or $\cup$?)\\
21	& 281.3757388	& --29.8844652	& 11.955	& 0.1122	& 1.519	&  9.245	& 13	& 	& $\sim^5$\\
22	& 284.4895524	& --30.1690194	& 13.343	& 0.0250	& 1.492	&  9.835	& 9	& 	& $\sim$\\
23	& 280.5309769	& --29.2564043	& 11.176	& 0.3118	& 1.483	&  9.095	& 13	& 	& $\rightarrow^5$\\
24	& 286.6381251	& --30.0461432	& 13.144	& 0.1182	& 1.479	& 10.587	& 11	& 	& $\cap$\\
25	& 281.7856196	& --29.9815077	& 14.321	& 0.1353	& 1.468	& 11.861	& 17	& 	& $\rightarrow$\\
26	& 283.6011999	& --30.4195844	& 13.399	& 0.0795	& 1.450	&  9.959	& 12	&C$^{ac}$& $\searrow$ or $\cup$\\
27	& 284.5604343	& --30.9197067	& 13.582	& 0.0254	& 1.432	& 10.461	& 12	& 	& $\cap$?\\
28	& 284.7247911	& --30.1433156	& 13.580	& 0.3951	& 1.400	&  9.539	& 12	& 	& $\searrow$\\
29	& 283.2143201	& --30.5832825	& 13.682	& 0.2242	& 1.347	& 10.201	& 12	& 	& $\cup$\\
30	& 282.8861938	& --29.7552222	& 13.643	& 0.0316	& 1.332	& 10.759	& 11	&C$^a$	& $\sim$\\
31	& 282.5503895	& --29.6363474	& 14.214	& 0.0623	& 1.299	& 11.208	& 12	& 	& $\sim$\\
\hline
\multicolumn{10}{p{0.95\textwidth}}{$^1$ Standard deviation of $Z$-band magnitudes. $^2$ Spectral types, where confirmed: C = carbon-rich; M = M-type (oxygen-rich); m/n = radial velocity member/non-member. References: a = Paper I; b = \citet{LZS+09}; c = \citet{WMIF99}; d = \citet{IG95}. $^3 \nearrow$: rising; $\searrow$: falling; $\cap$: apparent maxima; $\cup$: apparent minimum; $\sim$ undefined but long-period variation; $\rightarrow$ consistent with either noise or short-period variability. $^4$See notes in text. $^5$Saturated in $Z$-band.}\\
\hline
\end{tabular}
\end{minipage}
\end{table*}

\noindent
{\it RAGB 1} (IRAS F18568--3039, Sgr16): a very dusty object, as shown by ($K-[9]$), ($K-[11]$) and ($K-[12]$) colours from \citet{LZM+08} (hereafter LZM+08) and \emph{IRAS} of 5.0 to 5.4 mag. This makes it the dustiest star in the Lagadec et al.\ sample, with a dust-production rate of 1--$3 \times 10^{-8}$ M$_\odot$ yr$^{-1}$. \emph{WISE} data \citep{CWC+12} shows a slow redward increase in magnitude until at least 22 $\mu$m, consistent with carbon-rich dust.

\noindent
{\it RAGB 2:} though 2MASS \citep{CSvD+03} places ($J-K_{\rm s}$) at an even redder colour than ours, 4.63 mag, this star has not been extensively observed. \emph{WISE} indicates that this object may be dustier than RAGB 1, with [3.4]--[22] = 3.04 and 3.28 mag for RAGB 1 and RAGB 2, respectively.

\noindent
{\it RAGB 3} (IRAS 18556--3114): 2MASS shows this source to have been erroneously placed in the carbon-rich AGB as the saturated $J$ and $K_{\rm s}$ magnitudes are unreliable (2MASS shows $K_{\rm s} = 4.93$. \citet{RSM+12} indicates that this is a semi-regular variable with a 225-day period and an amplitude of $\delta V$ = 0.516 mag. The SuperWASP optical lightcurve \citep{BWA+10} broadly corroborates this, showing an amplitude of $\approx$0.9 mag in seven epochs over 300 days in their 400--700 nm filter range. This object is therefore most likely an intervening Bulge giant.

\noindent
{\it RAGB 4:} though we find little evidence of variation in our observations, literature data is consistent with strong variability. The 2MASS survey records magnitudes of $J,K_{\rm s}$ = 15.18 and 11.05 mag, some 2.0 and 1.1 magnitudes lower than ours, respectively. Meanwhile, DENIS records even lower magnitudes of $J,K_{\rm s}$ = 15.70 and 11.29 mag. WISE shows this star to be very dusty, with [3.4]--[22] = 3.11 mag. The data are therefore consistent with a highly variable, dusty star, with a long-period variation of $\delta K_{\rm s} \gtrsim 1.4$ mag, of which we have only observed its maximum.

\noindent
{\it RAGB 5} (IRAS F18413--3040, Sgr03): \citet{LZS+09} (hereafter LZS+09) present optical spectra, near-infrared photometry, and a mid-infrared \emph{Spitzer} IRS spectrum of this source. They show it to be a very dusty carbon star (optical depth $\tau_V = 9.56$) with a radial velocity which is low, but not inconsistent with galaxy membership, and estimate it to have a luminosity of 4529 L$_\odot$ for an estimated temperature of 2800 K. LZS+09 give the pulsation period to be 446 days, with a maximum around MJD 54\,340. Our observations (MJD 56\,034--56\,123) should therefore end close to maximum $K_{\rm s}$-band brightness. The fact that our light curve is still rising by MJD 56\,123 suggests a marginally longer period for this star, perhaps of 450--455 days. RAGB 5 is one of the most variable stars in our catalogue, showing an increase of $\delta Z$ = 1.3 mag over the 89 days of observation. For comparison, \citet{KY09} find variability of $\delta J > 1.11$ mag in 2MASS photometry, and LZS+09 give $\delta K = 1.15$ mag.

\noindent
{\it RAGB 6:} though one of the most variable among the reddened AGB stars ($\delta Z$ = 1.217 mag over 58 days), this star has not otherwise been extensively observed. Comparison to 2MASS and DENIS suggests $\delta J,K_{\rm s} > 2.73$ and 1.13 mag, respectively. This comparison also suggests that the $(J-K_{\rm s})$ colour could change significantly, from 1.8 mag in 2MASS to 3.0 mag here. This object is part of a four-star asterism of Sgr dSph giants, and is almost as optically bright as the nearby unreddened star. While optical measurements are sparse (the resolution of SuperWASP is not sufficient to separate the asterism's components), its optical counterpart is considerably brighter and bluer than would be expected from its $ZJHK_{\rm s}$ data. The lack of obvious variability in the optical suggests this is a binary companion, rather than the emergence of the central post-AGB star. An optical spectrum would be required to confirm this.

\noindent
{\it RAGB 7} (Sgr12, BD13-Sgr13a): LZM+08 and \emph{WISE} both indicate that this star is dusty. This matches the observations from \citep{BD13} (hereafter BD13), which show a rise of $\delta J = 1.0$ mag in three epochs covering our observation period (they also measure $\delta K = 0.5$ mag). However, we only observe a weakly rising light curve in $Z$, showing variability of $\delta Z = 0.36$ mag over the observing period. While amplitudes can change from cycle to cycle, the amplitude in $Z$ is very much smaller than at $J$-band. This could suggest an invariant hot component, such as a companion or other intervening star, which provides fractionally more flux at $Z$-band than $J$-band. A search for optical counterparts, however, has not revealed any obvious objects.

\noindent
{\it RAGB 8} (Sgr04): shows weak variation approaching a minimum between MJD 56\,055 and MJD 56,065. Comparison between VISTA and DENIS suggests $\delta K_{\rm s} > 0.67$ mag.

\noindent
{\it RAGB 9} (WMIF 15, Sgr17): \citet{WMIF99} note this star is a radial-velocity member and a carbon-rich Mira variable of approximate period 360 days, though they do not give any supporting information. We observe considerable variability with a minimum around MJD 56\,040. The star is optically faint (and red), and the corresponding SuperWASP lightcurve does not reflect this, suggesting a misassignment in their data.

\noindent
{\it RAGB 10} (Sgr09): both LZS+09 and Paper I identify this source as a carbon-rich radial-velocity member of the Sgr dSph, with negligible velocity shift between the two observations. The pulsation period was estiamted by LZS+09 at 370 days, placing the $K_{\rm s}$-band maximum shortly after the end of our observations (MJD 58\,082). The curvature of the lightcurve suggests a $Z$-band maximum is approaching in our final epoch, indicating a 360-day period might be more appropriate, depending on the phase lag between filters. LZS+09 model $\delta K_{\rm s} = 0.95$ mag, but our magnitude of $K_{\rm s} = 10.06 \pm 0.23$ mag is rather lower than their implied minimum of $K_{\rm s} = 9.55$ mag, implying a rather larger value. LZS+09 find the star to be luminous (6652 L$_\odot$) and dusty, but with a relatively low optical depth for this $(J-K_{\rm s})$ colour of 2.22.

\noindent
{\it RAGB 11} (Sgr08): minimum near MJD 56\,040. Comparison between 2MASS, DENIS and VISTA suggests relatively weak pulsation.

\noindent
{\it RAGB 12:} poorly studied, but with significant variability ($\delta Z = 1.01$ mag), plateauing or reaching a maximum sometime between MJD 56\,080 and 56\,125.

\noindent
{\it RAGB 13} (Sgr05): plateau or minimum sometime between MJD 56\,086 and 56\,123. Comparison between 2MASS, DENIS and VISTA suggests relatively weak pulsation.

\noindent
{\it RAGB 14} (WMIF17, Sgr19): \citet{WMIF99} classify this star as a semi-regular variable. Maximum probably reached around MJD 56\,090 to 56\,110.

\noindent
{\it RAGB 15} (C-2, WMIF 5, Sgr14, BD13-Sgr15): \citet{WMIF99} list this star as a Mira variable with period 228 days. Our observations cover 101 days, and are consistent with a minimum near MJD 56\,024 and a maximum after MJD 56\,125. BD13 broadly concur with these observations, finding $\delta K = 0.5$ mag and $\delta J = 1.0$ mag. We find $\delta Z > 1.0$ mag, but probably no more than $\delta Z \approx 1.1$ mag. Little phase lag can be present between $Z$ and $J$. \citet{IGI95} note that this is a radial-velocity member.

\noindent
{\it RAGB 16}: comparison of VISTA and 2MASS suggests the variability is rather larger than the $Z$-band data suggest: $\delta K_{\rm s} > 0.84$ mag. \citet{GSZ+10} find this an optically red star ($V-I$) = 3.38 mag.

\noindent
{\it RAGB 17}: maximum near MJD 56\,090.

\noindent
{\it RAGB 18}: this source has poor quality photometry due to being located near tile edges. A bluer colour of $(J-K_{\rm s}) = 1.1$--1.2 mag is suggested by DENIS and 2MASS. It is probably variable at the $\delta Z \sim 0.2$ mag level.

\noindent
{\it RAGB 19}: minimum near MJD 56\,060.

\noindent
{\it RAGB 20}: slow, shallow variability seen. Comparison to DENIS and 2MASS suggests this is probably the minimum of a light curve with a photometric range of $\delta K \gtrsim 0.4$ mag.

\noindent
{\it RAGB 21}: though the data is slightly saturated, a probable rise in magnitude over the observing period can be inferred.

\noindent
{\it RAGB 22}: though the variability observed here is of very low amplitude, comparison of the DENIS and 2MASS magnitudes suggest that we observe a minimum of long-term variability which is likely to be of much greater photometric range, perhaps $\delta K_{\rm s} = 0.79$ mags or even more.

\noindent
{\it RAGB 23}: as with RAGB 3, this saturated star is rather brighter than indicated here, with $K_{\rm s} = 7.96$ in 2MASS.

\noindent
{\it RAGB 24}: maximum near MJD 56\,080. The observations are consistent with a period of well over 100 days.

\noindent
{\it RAGB 25}: a fainter object, where variability is difficult to accurately determine due to its placement on the overlap region between tiles, though variability is expected to be slight. VISTA, 2MASS and DENIS consistently find a comparatively faint $K_{\rm s}$-band magnitude of around 11.7 mag. \emph{WISE} photometry is consistent with there being no dust excess. The object is not obviously optically resolved and may therefore represent a foreground early M-type dwarf.

\noindent
{\it RAGB 26} (WMIF 6, UKST 12, Sgr13): \citet{WMIF99} note that this star is a carbon-rich, 280-day semi-regular variable. \citet{WIC96} find 0.16 mag of variability in $K_{\rm s}$ between two epochs. LZS+09 note that this star is relatively blue, but still has nearly a magnitude of 9-$\mu$m excess. This is corroborated by the \emph{WISE} photometry. This source was observed by \emph{ISO} in 1997, but was only detected at reasonable signal-to-noise at 4.5 $\mu$m: photometry which has been superceded by \emph{WISE}. The lightcurve shows a decline of 0.3 mags, with a possible minimum between MJD 56\,080 and 56\,120. 

\noindent
{\it RAGB 27} (WMIF 11, UKST 18): quoting this star as a radial-velocity member, \citet{WIC96} also note some slight variability in $K_{\rm s}$, which they quote in \citet{WMIF99} as being semi-regular. \emph{WISE} data shows little dust emission, though the 22 $\mu$m data are of poor signal-to-noise. This source was also observed by \emph{ISO}. Variation between VISTA, 2MASS and DENIS data suggest the variability of this source is relatively slight, and that the $\sim$0.1-mag variability in $Z$ we observe probably represents most of the lightcurve.

\noindent
{\it RAGB 28:} \emph{WISE} shows considerable dust production in this strongly varying source.

\noindent
{\it RAGB 29} (BD13-Sgr12): BD13 find a $J$-band minimum near MJD 56\,080, as we find in $Z$. BD13 also find an earlier minimum, which they use to ascribe a 210-day period to this star. They find $\delta K = 0.4$ mag and $\delta J = 1.0$ mags. We find $\delta Z > 0.66$ mag and, on the basis of the light curve presented by BD13, suggest $\delta Z \approx 1.1$ mag. \emph{WISE} magitudes are consistent with cold carbon-rich or warmer oxygen-rich dust.


\noindent
{\it RAGB 30:} weak variability is seen in this star, with a maximum reached at MJD 56\,064 followed by a minimum at MJD 56\,078. We suggest this star is a low-amplitude semi-regular variable. Shown in Paper I to be carbon-rich.



\noindent
{\it RAGB 31:} a rather fainter object, with variability which may repeat with an 60-day period and a photometric range in $Z$ of 0.2--0.3 mag.

\subsection{Unreddened AGB stars}


\begin{table*}
 \centering
 \begin{minipage}{160mm}
  \caption{Variability of surveyed un-reddened AGB stars. Footnotes are as in Table \ref{RAGBTable}. }
\label{UAGBTable}
  \begin{tabular}{rlllllrrcc}
  \hline\hline
   \multicolumn{1}{c}{UAGB}	&    \multicolumn{1}{c}{RA}	&    \multicolumn{1}{c}{Dec}
& \multicolumn{1}{c}{$Z$}	& \multicolumn{1}{c}{$\sigma Z^{(1)}$}	
& \multicolumn{1}{c}{($J-K_{\rm s}$)}	& \multicolumn{1}{c}{$K_{\rm s}$}	
& \multicolumn{1}{c}{$n_{\rm epochs}$}	& \multicolumn{1}{c}{Spectral}
& \multicolumn{1}{c}{Lightcurve}\\
   \multicolumn{1}{c}{ID}	&    \multicolumn{1}{c}{Degrees}&    \multicolumn{1}{c}{Degrees}&
\multicolumn{1}{c}{(mag)} & \multicolumn{1}{c}{(mag)} & \multicolumn{1}{c}{(mag)} & \multicolumn{1}{c}{(mag)} & \multicolumn{1}{c}{\ } & \multicolumn{1}{c}{type$^{2}$} & \multicolumn{1}{c}{shape$^{3}$} \\
 \hline
 1 & 280.5976609 & -29.6720736 & 11.4044 & 0.5550 & 1.293 &  9.361 & 14 &n?$^5$& $\rightarrow^5$\\
 2 & 281.6630154 & -30.7646466 & 12.7089 & 0.1818 & 1.271 &  9.546 & 13 &Cm$^c$& $\searrow$ or $\cap$\\
 3 & 280.5218739 & -29.5665088 & 11.6405 & 0.1836 & 1.356 &  9.555 & 13 &n?$^5$& $\rightarrow^5$\\
 4 & 284.8218384 & -31.3447208 & 13.0747 & 0.1428 & 1.271 &  9.742 & 13 &Mm$^a$& $\cup$ \\
 5 & 286.6156416 & -30.9516256 & 13.0208 & 0.0447 & 1.261 &  9.824 & 12 &      & $\sim$\\
 6 & 280.5361679 & -29.6961108 & 12.1821 & 0.0331 & 1.184 &  9.825 & 11 &n?$^5$& $\rightarrow^5$\\
 7 & 285.1181780 & -30.6241526 & 12.9145 & 0.0652 & 1.224 &  9.841 & 11 &Mm$^a$& $\cap$\\
 8 & 284.6516836 & -31.3396958 & 13.1570 & 0.0919 & 1.266 &  9.889 & 13 &Mm$^a$& $\cup$\\
 9 & 285.2923234 & -30.0774220 & 11.1311 & 0.2905 & 1.240 &  9.943 & 12 &n?$^5$& $\rightarrow^5$\\
10 & 280.6646219 & -29.5914609 & 12.6666 & 0.0186 & 1.339 &  9.990 & 13 &      & $\rightarrow$\\
11 & 280.1585944 & -29.8341750 & 13.0754 & 0.0433 & 1.175 & 10.009 & 13 &      & $\cap$ or $\sim$\\
12 & 281.1450978 & -31.4358818 & 13.0211 & 0.0198 & 1.193 & 10.022 & 12 &n?$^5$& $\rightarrow$ or $\searrow$\\
13 & 280.8483272 & -30.4010066 & 11.4006 & 0.1576 & 1.224 & 10.023 & 11 &      & $\rightarrow^5$\\
14 & 283.3153455 & -30.2992199 & 13.1276 & 0.0576 & 1.236 & 10.031 & 24 &Mm$^a$& $\cup$\\
15 & 281.6670318 & -29.5231793 & 13.0288 & 0.0994 & 1.173 & 10.075 & 13 &      & $\cup$ or $\sim$\\
16 & 283.3085278 & -30.3205402 & 12.8843 & 0.0306 & 1.194 & 10.079 & 24 &Mn$^a$& $\rightarrow$ or $\sim$\\
17 & 284.8465122 & -30.1054795 & 13.2374 & 0.0362 & 1.308 & 10.082 & 12 &      & $\sim$\\
18 & 285.1227237 & -31.0537430 & 13.0260 & 0.0821 & 1.146 & 10.082 & 12 &Mm$^a$& $\cap$\\
19 & 284.7709701 & -30.0975437 & 12.7917 & 0.0206 & 1.143 & 10.083 & 12 &      & $\sim$\\
20 & 281.8735951 & -29.9628605 & 13.0052 & 0.0137 & 1.179 & 10.094 & 11 &      & $\rightarrow$\\
21 & 286.5770099 & -30.0554471 & 13.1746 & 0.0833 & 1.274 & 10.125 & 12 &      & $\sim$\\
22 & 280.3223572 & -29.7626700 & 13.3534 & 0.1575 & 1.202 & 10.137 & 13 &      & $\cup$\\
23 & 280.2324429 & -31.2846753 & 12.9928 & 0.1451 & 1.187 & 10.137 & 13 &      & $\nearrow$ or $\cap$\\
24 & 283.9335039 & -30.9745404 & 13.2781 & 0.0428 & 1.138 & 10.144 & 12 &      & $\cup$ or $\sim$\\
25 & 283.8990004 & -31.0187357 & 12.9393 & 0.0986 & 1.164 & 10.147 & 11 &      & $\cap$\\
26 & 280.5304039 & -30.9248461 & 13.3442 & 0.0475 & 1.232 & 10.203 & 13 &      & $\searrow$ or $\cup$\\
27 & 281.3931408 & -31.2490268 & 13.2115 & 0.0391 & 1.220 & 10.215 & 13 &      & $\sim$\\
28 & 284.8388035 & -30.3644644 & 13.1554 & 0.0201 & 1.258 & 10.238 & 12 &      & $\rightarrow$\\
29 & 284.9541629 & -31.1671258 & 13.3251 & 0.1849 & 1.171 & 10.294 & 25 &Mm$^a$& $\sim$ or $\rightarrow$\\
30 & 284.6926108 & -31.5116994 & 13.4482 & 0.3245 & 1.145 & 10.306 & 13 &Mm$^a$& $\cup$\\
31 & 284.0856013 & -30.3825261 & 13.1949 & 0.0217 & 1.154 & 10.321 & 12 &Mm$^a$& $\sim$\\
32 & 281.6060343 & -29.7380383 & 13.3004 & 0.0194 & 1.201 & 10.339 & 13 &      & $\rightarrow$\\
33 & 282.8673956 & -29.9884093 & 13.1122 & 0.0723 & 1.188 & 10.348 & 12 &Mm$^a$& $\cap$ or $\sim$\\
34 & 283.1796691 & -29.2678719 & 13.2841 & 0.1107 & 1.178 & 10.348 & 12 &      & $\cup$\\
35 & 286.0356460 & -31.0563244 & 13.3103 & 0.0437 & 1.147 & 10.348 & 12 &      & $\sim$\\
36 & 281.0002121 & -30.5574332 & 13.1602 & 0.0179 & 1.211 & 10.349 & 11 &      & $\rightarrow$ or $\sim$\\
37 & 281.2569201 & -29.2128311 & 13.2176 & 0.0146 & 1.216 & 10.359 &  9 &      & $\rightarrow$\\
38 & 286.3362454 & -30.2777762 & 13.1809 & 0.0488 & 1.172 & 10.374 & 12 &      & $\sim$\\
39 & 286.1855546 & -30.0910232 & 12.8398 & 0.0400 & 1.120 & 10.380 & 11 &      & $\cap$\\
40 & 286.3625798 & -30.1227183 & 13.1248 & 0.0326 & 1.163 & 10.397 & 11 &      & $\sim$\\
41 & 284.9086093 & -30.9539019 & 13.0894 & 0.1079 & 1.117 & 10.401 & 12 &Mm$^a$& $\cap$\\
42 & 280.8162354 & -31.0641027 & 13.2282 & 0.0393 & 1.211 & 10.418 & 13 &      & $\sim$\\
43 & 281.6784795 & -29.7488860 & 13.3112 & 0.0248 & 1.190 & 10.424 & 13 &      & $\sim$\\
44 & 282.6389918 & -30.2390686 & 13.2992 & 0.0247 & 1.178 & 10.428 & 12 &      & $\sim$\\
45 & 284.9913406 & -31.7411886 & 13.2588 & 0.0501 & 1.301 & 10.439 & 12 &      & $\sim$\\
46 & 285.5398350 & -30.9428496 & 13.2787 & 0.0980 & 1.139 & 10.443 & 12 &      & $\sim$\\
47 & 280.9562777 & -31.1001157 & 13.3448 & 0.0044 & 1.204 & 10.454 &  9 &      & $\rightarrow$\\
48 & 281.2061561 & -31.3288772 & 13.2276 & 0.0459 & 1.144 & 10.456 & 13 &      & $\sim$\\
49 & 286.0141906 & -30.0693296 & 13.1508 & 0.0467 & 1.109 & 10.461 & 12 &      & $\cup$ or $\sim$\\
50 & 282.7942289 & -30.9299523 & 13.2760 & 0.0624 & 1.241 & 10.466 & 12 &      & $\sim$\\
51 & 281.8295341 & -30.7626729 & 13.3967 & 0.0273 & 1.205 & 10.467 & 12 &      & $\sim$\\
52 & 280.6488960 & -30.3108298 & 13.9577 & 0.7060 & 1.246 & 10.470 & 26 &      & $\nearrow$ or $\cap$\\
53 & 280.1729928 & -31.3064766 & 13.1175 & 0.0174 & 1.125 & 10.477 & 13 &      & $\rightarrow$ or $\sim$\\
54 & 283.4837779 & -31.6584513 & 13.1875 & 0.0460 & 1.125 & 10.487 & 13 &      & $\sim$\\
55 & 283.2848249 & -30.7127248 & 13.3436 & 0.0134 & 1.205 & 10.494 & 12 &Mm$^a$& $\rightarrow$ or $\sim$\\
56 & 280.7829580 & -31.3370932 & 13.2248 & 0.1195 & 1.152 & 10.508 & 13 &      & $\cup$\\
57 & 282.8896894 & -29.3956040 & 13.2868 & 0.1205 & 1.186 & 10.508 & 12 &      & $\sim$\\
58 & 281.1340550 & -31.3355882 & 13.2635 & 0.0322 & 1.177 & 10.509 & 13 &      & $\sim$\\
59 & 282.9903638 & -29.2164024 & 13.2069 & 0.0094 & 1.147 & 10.515 & 12 &      & $\rightarrow$\\
60 & 280.1189109 & -30.3164445 & 13.2874 & 0.0176 & 1.167 & 10.530 & 26 &      & $\rightarrow$ or $\nearrow$\\
61 & 281.3691555 & -30.9807642 & 13.4161 & 0.0539 & 1.178 & 10.541 & 13 &      & $\cup$\\
62 & 286.6485045 & -30.4580491 & 13.3033 & 0.0252 & 1.166 & 10.548 & 12 &      & $\sim$ or $\cup$\\
\hline
\end{tabular}
\end{minipage}
\end{table*}

\noindent
{\it UAGB 1, 3, 6, 12:} bright sources with saturated $J$ and $K_{\rm s}$ magnitudes, which in 2MASS are consistent with Bulge giants. UAGB12 is near the Bulge RGB tip, with unsaturated $Z$-band magnitudes showing a weak decline in brightness.

\noindent
{\it UAGB 2} (UKST 3, WMIF 2, Sgr06): \citet{WMIF99} report this as a carbon-rich member, with a 300-day Mira-amplitude variability. Our lightcurve suggests a period of $\gg$100 days which has $\delta Z > 0.7$ mag, and a $Z$-band maximum around MJD 56\,040.

\noindent
{\it UAGB 4:} minimum observed near MJD 56\,110, possible maximum near MJD 56\,020, suggesting $\delta Z \sim 0.5$ mag with period $\sim$180 days. Confirmed M-giant member in Paper I.

\noindent
{\it UAGB 5} (V1246 Sgr, ASAS 190629--3055.1): discoverSUK+04,TSI13ed by \citet{vanHouten53}, who gives an ephemeris of MJD 26\,227.5 + 125 $e$ and a photographic range of $\delta p > 2.4$ mag. This ephemeris predicts a maximum at MJD 56\,102.5, close to that observed. After 239 cycles, however, an error of only 12 hours corresponds to a full period. We observe a minimum near MJD 56\,048 and a maximum sometime between MJD 56\,083 and MJD 56\,141. The ASAS lightcurve \citep{Pojmanski97} is incomplete, as the star is too faint at minimum. However, ASAS records $\delta V > 2.8$ mag, with maxima generally declining over their observing period, suggesting an addition long, secondary period. A combination of the original ephemeris, the ASAS lightcurve and our data can be fit with any period between 124.8 and 128.3 days, with 126.6 days appearing the most likely solution.

\noindent
{\it UAGB 8:} minimum observed near MJD 56\,080, possibly bounded by two maxima near MJD 56\,020 and 56\,140 (little rise is seen between MJD 56\,123 and 56\,141), suggesting $\delta Z \sim 0.3$ mag with period $\sim$100--120 days. Confirmed M-giant member in Paper I.

\noindent
{\it UAGB 9:} a bright source with saturated $J$ and $K_{\rm s}$ magnitudes, which in 2MASS are consistent with a foreground F star.

\noindent
{\it UAGB 10:} 2MASS gives a much lower $K_{\rm s} = 10.22$ magnitude, hence also $(J-K_{\rm s}) = 1.04$ mag colour. These colours would place it between the Sgr dSph and Bulge populations. It is not clear why the difference arises. \emph{WISE} magnitudes are more consistent with the 2MASS magnitude.

\noindent
{\it UAGB 16:} confirmed M-giant non-member with near-solar velocity in Paper I. Little long-term variation is seen. The red colour ($(J-K_{\rm s}) = 1.20$ mag) of 2MASS and DENIS are confirmed. The lack of variability argues against an M-giant member with a peculiar velocity. There is no indication in the spectrum of a binary companion, leading us to conclude that it is an unusually red Bulge star.

\noindent
{\it UAGB 20:} an object on the blue boundary of the colour-selected AGB, exceptional for its pronounced lack of variability. Nevertheless, a small rise of $\delta Z \approx 0.03$ mag can be seen around MJD 56\,080, and we suggest a period of $\sim$80 days or longer.

\noindent
{\it UAGB 21:} a minimum and maximum are observed near MJD 56\,080 and 56\,080, respectively. The lightcurve suggests either a 80-day symmetrical or 120-day asymmetrical periodicity.

\noindent
{\it UAGB 23:} an unusual lightcurve, showing a rise of $\delta Z = 0.4$ mag to a constant level, as in an R CrB star or eclipsing binary exiting a minimum. Assuming it is a pulsating variable suggests a lightcurve that deviates significantly from sinusoidal. \emph{WISE} shows the star to be considerably dusty.

\noindent
{\it UAGB 27:} maxima are observed near MJD 56\,035 and 56\,080 and a minimum near MJD 56\,057. The single epoch at MJD 56\,123 has a magnitude consistent with maximum light. This suggests a sinusoidal lightcurve with a period of $\sim$45 days and photometric range of $\delta Z = 0.11$ mag. The star appears dusty in \emph{WISE} with colours consistent with silicate dust.

\noindent
{\it UAGB 29:} this object lies on a tile boundary containing the under-performing VISTA chip. Results from the fully functioning chip suggest little variation, only $\delta Z \approx 0.064$ mag. The star is a known M-giant member (Paper I). The SuperWASP light curve contains six nights of observations which do not strongly constrain any variability. \emph{WISE} shows the star to be considerably dusty, with colour consistent with silicate dust.

\noindent
{\it UAGB 30:} this M-giant member shows marked variability of $\delta Z > 1.1 mag$, including a minimum near MJD 56\,050.

\noindent
{\it UAGB 31:} relatively little but noticable variability ($\delta Z \approx 0.1 mag$) seen over timescales of $\sim$40 days in this M-giant member, though not enough to constrain a period.

\noindent
{\it UAGB 33:} light curve suggests period close to the coverage period of 101 days, with $\delta Z \approx 0.25$ mag. M-giant member.

\noindent
{\it UAGB 35:} light curve suggests period of 70--80 days with $\delta Z \approx 0.15$ mag.

\noindent
{\it UAGB 42:} period appears to be $\sim$60 days with $\delta Z \approx 0.15$ mag, but with a longer, secondary period.

\noindent
{\it UAGB 45} (BD13-Sgr15): BD13 note this star to be a low-amplitude irregular variable, with $\delta K = 0.8$ mag and $\delta J = 0.6$ mag. Our observations are consistent with these findings, as the later epochs do not match the variability in the earlier epochs. We find $\delta Z > 0.2$ mag.

\noindent
{\it UAGB 46:} period appears to be close to 100 days, with $\delta Z = 0.35$ mag.

\noindent
{\it UAGB 48:} possible period of $\approx$80--90 days.

\noindent
{\it UAGB 50:} possible period of $\approx$80 days.

\noindent
{\it UAGB 52:} marked variability of $\delta Z > 2.3$ mag, over a period of $>$100 days. Comparison with 2MASS photometry suggests $\delta K > 1.0$ mag. \emph{WISE} photometry are consistent with a very dusty star, with [3.4]--[22] = 3.0 mag. The VISTA, 2MASS and \emph{WISE} colours imply an oxygen-rich star with predominantly silicate dust.

\subsection{Blue AGB stars}


\begin{table*}
 \centering
 \begin{minipage}{160mm}
  \caption{Candidate variables detected among candidate blue AGB stars. Footnotes are as in Table \ref{RAGBTable}. }
\label{BAGBTable}
  \begin{tabular}{rlllllrrcc}
  \hline\hline
   \multicolumn{1}{c}{BAGB}	&    \multicolumn{1}{c}{RA}	&    \multicolumn{1}{c}{Dec}
& \multicolumn{1}{c}{$Z$}	& \multicolumn{1}{c}{$\sigma Z^{(1)}$}	
& \multicolumn{1}{c}{($J-K_{\rm s}$)}	& \multicolumn{1}{c}{$K_{\rm s}$}	
& \multicolumn{1}{c}{$n_{\rm epochs}$}	& \multicolumn{1}{c}{Spectral}
& \multicolumn{1}{c}{Lightcurve}\\
   \multicolumn{1}{c}{ID}	&    \multicolumn{1}{c}{Degrees}&    \multicolumn{1}{c}{Degrees}&
\multicolumn{1}{c}{(mag)} & \multicolumn{1}{c}{(mag)} & \multicolumn{1}{c}{(mag)} & \multicolumn{1}{c}{(mag)} & \multicolumn{1}{c}{\ } & \multicolumn{1}{c}{type$^{2}$} & \multicolumn{1}{c}{shape$^{3}$} \\
 \hline
 11 & 282.9353140 & -29.4197788 & 12.8241 & 0.1120 & 1.073 &  9.900 & 12 &      & $\cap$\\
 12 & 281.4217715 & -31.0279792 & 13.0167 & 0.1419 & 1.140 &  9.905 & 13 &      & $\cap$ or $\sim$\\
 13 & 285.0473668 & -30.3345838 & 13.3651 & 0.2004 & 1.096 &  9.929 & 12 &      & $\cup$\\
 15 & 280.1516272 & -30.1587416 & 12.6358 & 0.0468 & 1.120 &  9.951 & 13 &      & $\sim$\\
 17 & 284.2326725 & -31.4113585 & 12.9832 & 0.0400 & 1.096 &  9.982 & 13 &Cm$^c$& $\sim$\\
 18 & 281.7870426 & -31.1333227 & 12.9803 & 0.0650 & 1.091 & 10.017 & 17 &      & $\searrow$ or $\cup$\\
 19 & 282.6839956 & -29.3966201 & 12.9683 & 0.0786 & 1.115 & 10.039 & 12 &      & $\sim$\\
 22 & 285.6400721 & -30.9351049 & 13.0315 & 0.0443 & 1.089 & 10.051 & 12 &      & $\sim$\\
 23 & 283.2095889 & -29.9421378 & 13.0033 & 0.0397 & 1.069 & 10.056 & 12 &Cm$^{ac}$& $\searrow$ or $\sim$\\
 26 & 281.2284190 & -29.8451271 & 13.1144 & 0.0648 & 1.058 & 10.074 & 13 &      & $\sim$\\
 29 & 281.1256669 & -30.3424262 & 13.1112 & 0.0302 & 1.064 & 10.100 & 26 &M?m$^d$& $\cup$\\
 30 & 283.8756147 & -31.0593310 & 12.7524 & 0.0247 & 1.070 & 10.130 & 12 &      & $\sim$\\
 32 & 284.9963922 & -30.7007016 & 13.2079 & 0.1936 & 1.114 & 10.177 & 12 &      & $\cup$ or $\sim$\\
 34 & 283.4033461 & -29.7365818 & 13.3076 & 0.0613 & 1.108 & 10.194 & 12 &      & $\searrow$ or $\cup$\\
 36 & 284.9056321 & -31.7915792 & 12.9673 & 0.2238 & 1.025 & 10.256 & 13 &      & $\nearrow$ or $\cap$\\
 39 & 286.2368038 & -30.9730329 & 12.9957 & 0.0847 & 1.054 & 10.266 & 12 &      & $\sim$\\
 43 & 282.8209459 & -30.4854181 & 13.1346 & 0.1761 & 1.054 & 10.341 & 12 &      & $\searrow$\\
 44 & 280.4763145 & -29.2010908 & 13.2414 & 0.0411 & 1.054 & 10.350 & 13 &      & $\cup$\\
 46 & 285.3682748 & -30.2029746 & 12.8466 & 0.0805 & 1.075 & 10.351 & 12 &      & $\cap$\\
 55 & 282.7055679 & -30.1344021 & 12.9756 & 0.0874 & 1.055 & 10.409 & 12 &      & $\searrow$ or $\cap$\\
 59 & 284.0777288 & -32.2264021 & 12.7961 & 0.1144 & 1.042 & 10.433 & 13 &      & $\cap$\\
 67 & 282.5152271 & -29.8846736 & 13.2412 & 0.0281 & 1.007 & 10.510 & 12 &M$^a$ & $\cup$ or $\sim$\\
 70 & 283.3254966 & -30.2287857 & 13.0622 & 0.0603 & 1.062 & 10.517 & 24 &Mm$^a$& $\sim$\\
 76 & 282.1601709 & -29.3217070 & 13.3088 & 0.0321 & 1.049 & 10.540 & 12 &      & $\sim$\\
 82 & 282.0554496 & -30.3151355 & 13.1574 & 0.1270 & 1.065 & 10.583 & 22 &      & $\cap$\\
 98 & 285.5233887 & -30.1867587 & 13.3532 & 0.0335 & 1.043 & 10.660 & 12 &      & $\sim$\\
101 & 285.2764826 & -30.1839422 & 12.8300 & 0.0651 & 1.029 & 10.674 & 10 &      & $\cap$\\
104 & 282.2274877 & -29.5043855 & 13.3038 & 0.0481 & 1.048 & 10.685 & 12 &      & $\sim$\\
108 & 284.4264517 & -31.4839290 & 13.2648 & 0.0677 & 0.989 & 10.697 & 13 &      & $\sim$\\
109 & 284.1534754 & -31.3480006 & 13.0566 & 0.1076 & 0.995 & 10.697 & 11 &      & $\cap$\\
124 & 284.8466210 & -30.0982754 & 13.2834 & 0.0382 & 1.053 & 10.739 & 12 &      & $\sim$\\
132 & 280.9164152 & -30.1373089 & 13.3573 & 0.2407 & 1.022 & 10.777 & 13 &      & $\sim$\\
\hline
\end{tabular}
\end{minipage}
\end{table*}

\noindent
{\it BAGB 11:} reaches maximum near MJD 56\,090 with period $\gg$100 days and $\delta Z > 0.35$ mag.

{\it BAGB 12:} probable minimum near MJD 56\,040 and apparent maximum between MJD 56\,090 and 56\,100. Period is probably $\sim$150 days with $\delta Z \approx 0.4$ mag. \emph{WISE} indicates this star has silicate-rich dust.

{\it BAGB 13:} minimum observed on or near MJD 56\,074. Variation of $\delta Z > 0.65$ mag with period $\gg$100 days.

{\it BAGB 15:} small-amplitude variable ($\delta Z \approx 0.15$ mag) with minimum observed near MJD 56\,050 and maxmimum near MJD 56\,068. Period is likely to be $\sim$50--60 days with an additional long secondary period.

{\it BAGB 17} (UKST 15, WMIF 8): a small-amplitude variable of $\delta Z \approx 0.15$ mag with period $\sim$120 days. WMIF note this star as a semi-regular variable. A much larger colour of $(J-K_{\rm s}) = 1.53$ mag is given by 2MASS, and 1.56 mag by DENIS. Narrowly missed by our automated variable finder.

{\it BAGB 18:} shows a decline by $\delta Z = 0.25$ mag over the observation period, with a possible upturn by MJD 56\,123.

{\it BAGB 19:} shows a decline by $\delta Z = 0.35$ mag over the period MJD 56\,046 -- 56\,082. Outlying points at MJD 56\,024 and MJD 56\,125 suggest a period of $\sim$100 days.

{\it BAGB 22:} a small but erratic decline in magnitude is seen.

{\it BAGB 23} (C3, WMIF 4, Sgr11): possible maxmimum near MJD 56\,040. WMIF note this star is a semi-regular variable. We find $\delta Z > 0.14$ mag, with period $\gg$100 days.

{\it BAGB 26:} minimum observed near MJD 56\,050; $\delta Z > 0.20$ mag.

{\it BAGB 29} (IG95 400-04): a radial-velocity member in \citet{IG95}. We presume this is a M-type star, as we would expect Ibata \& Gilmore to otherwise note its carbon-rich nature \citet{IGI95}. We find $\delta Z > 0.11$ mag, but is probably not much greater than this, as the photometry suggets a period of a little over the observing period of 100 days.

{\it BAGB 30:} minimum observed near MJD 56\,045; maximum observed near MJD 56\,075. A period of $\approx$80 days is suggested, with photometric range $\delta Z \approx 0.08$ mag. Not classed as variable by our automated variable finder.

{\it BAGB 32} (BD13-Sgr16): BD13 note a 299-day period with $\delta K = 0.5$ mag and $\delta J = 1.1$ mag. We find $\delta Z > 0.55$ mag, and observe a minimum near MJD 56\,050, suggesting little phase lag between $J$ and $Z$. BD13 also note a much larger $(J-K)$ colour of 1.56 mag, which is closer to the 2MASS and DENIS colours of $(J-K_{\rm s})$ = 1.61 and 1.66 mag, respectively.

{\it BAGB 34:} decline observed, with probable minimum shortly after MJD 56\,080.

{\it BAGB 36:} sharp rise of $\delta Z \approx 0.6$ mag seen to MJD 56\,100 followed by a possible slow decline suggestive of a period of $\gg$100 days.

{\it BAGB 39:} shows a decline by $\delta Z = 0.25$ mag over the period MJD 56\,046 -- 56\,082. Outlying points at MJD 56\,024 and MJD 56\,141 suggest a period of $\sim$120 days.

{\it BAGB 43:} consistent decline over 101 days showing variablity of $\delta Z > 0.75$ mag. Small modulations on this lightcurve can be seen, but a periodicity of $\gg$100 days is inferred.

{\it BAGB 44:} variations of $\gg$100 days and $\delta Z > 0.15$ mag seen, though probably of relatively small amplitude. Minimum observed near MJD 56\,060.

{\it BAGB 46:} maximum observed on or near MJD 56\,074; the period is $\gg$120 days with $\delta Z > 0.30$ mag.

{\it BAGB 55:} maximum observed near MJD 56\,045; the period is $\gg$100 days with $\delta Z > 0.30$ mag.

{\it BAGB 59:} maximum observed on or near MJD 56\,078; the period is $\gg$120 days with $\delta Z > 0.40$ mag.

{\it BAGB 67:} minimum obsreved near MJD 56\,065. A low-amplitude variable but with $\delta Z > 0.08$ mag, the period is likely to be slightly longer than the 100-day observing window.

{\it BAGB 70:} probable minimum obsreved near MJD 56\,040 and maximum near MJD 56\,080. This star has a suggested variability of $\delta Z \approx 0.20$ mag, and the period is likely to be slightly longer than the 100-day observing window.

{\it BAGB 76:} a small-amplitude variable with a period of a little under 100 days with $\delta Z \sim 0.1$ mag. A maximum is observed near MJD 56\,075. There may be a long, secondary period. Not classed as variable by our automated variable finder.

{\it BAGB 82:} the photometric range is $\delta Z > 0.45$ mag and the period is $\gg$60 days. A maxmimum is observed near MJD 56\,068.

{\it BAGB 98:} a small-amplitude variable with $\delta Z = 0.08$ or 0.09 mag. A period of 60--70 days is inferred from the light curve. Narrowly missed by our automated variable finder.

{\it BAGB 101:} a maximum is observed near MJD 56\,074, with $\delta Z \approx 0.20$ mag over the short 36-day observing period.

{\it BAGB 104:} a mimimum is observed near MJD 56\,070, and a potential period identified of $\approx$70 days, with $\delta Z \approx 0.15$ mag.

{\it BAGB 108:} a complex lightcurve suggesting more than one puslation mode is excited. We observe $\delta Z = 0.22$ mag over the 120-day observing period.

{\it BAGB 109:} a maximum is observed near MJD 56\,075, indicative of a period of $\gg$60 days with $\delta Z > 0.4$ mag.

{\it BAGB 124:} another complex lightcurve suggesting more than one puslation mode is excited. We observe $\delta Z = 0.14$ mag over the 100-day observing period.

{\it BAGB 132:} a smooth lightcurve with a mimimum very close to MJD 56\,050 and a maxmimum near MJD 56\,090, giving $\delta Z \approx 0.63$ mag. We suggest a period of little more than 100 days.


\label{lastpage}

\end{document}